\documentclass[aps,prb,preprint,eqsecnum,floatfix]{revtex4-1}

\usepackage{graphicx,epstopdf}
\usepackage{bm,bbm}
\usepackage{hyperref}
\usepackage{color}
\usepackage{subfigure}
\usepackage{amsmath}
\usepackage{amssymb}
\usepackage{float}
\def\rvec{{\bf r}}
\def\kvec{{\bf k}}
\def\pvec{{\bf p}}
\def\qvec{{\bf q}}
\def\bra#1{\bigl\langle{ #1} \bigr|}
\def\ket#1{\bigl|{ #1} \bigr\rangle}

\def\ovlp#1#2{\bigl\langle{ #1}\big|{#2} \bigr\rangle}

\def\hm#1#2{\frac{\hbar^2}{ #1m_{#2}}}
\def\he#1{$^{#1}$He}
\def\half{\frac{1}{2}}

\def\Re{{\cal R}e}
\def\Im{{\cal I}m}
\def\I{{\rm i}}
\def\etal{{\em et al.\/}\ }
\def\ie{{\em i.e.\/}\ }

\def\boxit#1{
        \centerline{\vbox{\hsize=6.0truein\hrule\hbox{\vrule\kern5pt
        \vbox{\kern5pt\noindent #1\smallskip
        \kern5pt}\kern5pt\vrule}\hrule}
}}
\def\he#1{$^{#1}$He}

\begin{document}
%\tightenlines

\title{Dynamic Many-Body Theory: The dynamics of atomic impurities in $^4$He}
\author{E.~Krotscheck}
\affiliation{Department of Physics, University at Buffalo, SUNY
Buffalo NY 14260}

\affiliation{Institut f\"ur Theoretische Physik, Johannes
Kepler Universit\"at, A 4040 Linz, Austria}

\begin{abstract}
%%%%%%%%%%%%%%%%%%%%%%%%%%%%%
We implement manifestly microscopic many-body methods to study the
dynamics of atomic impurities in a host quantum fluid, specifically
\he4.  Our investigations are motivated by experiments of muonium
atoms within \he4 with the goal of testing the universality of free
fall by neutral bound states using unstable particles.

Structure calculations are performed using standard semi-analytic
methods; we extend here the calculations of our previos work (Journal
of Low Temperature Physics {\bf 93}, 415-449 (1993)) to muonic \he4,
antiprotonic \he4 and mounium atoms within \he4. We find that the
muonium impurity probes, due to its large zero-point motion, the
atomic interaction at rather short distances. Its chemical potential
is estimated to be about 19 meV.  Antiprotonic \he4 has, on the other
hand, a negative chemical potential.

Dynamics is treated by making all correlation functions
time-dependent. In analogy to the derivation of the dynamics of the
background liquid, we derive a working formula for the impurity
self-energy that includes the most relevant physical effects.  Results
for the effective mass of \he3 and hydrogen atoms agree well with
available experiments. The dispersion relations of muonic \he4 and
antiprotonic \he4 pass through under the roton minimum.
\end{abstract}
\maketitle

  \section{Introduction}

Recent experiments \cite{Muonium} have reported the synthesis of a
high-brightness muonium beam, extracted from superfluid helium by
exploiting its chemical potential and transport properties. The method
uses a thin layer of superfluid \he4 at a temperature near 0.3 K; the
purpose of these experiments to test the univerality of free fall for
unstable particles. In view of these experiments we have extended our
previous \cite{SKJLTP} studies on hydrogen isotopes to muonium and
similar atoms.  Hydrogen impurities have a positive chemical potential
which has been measured for deuterium \cite{Hayden95}. Our
calculations for muonium give a chemical potential of about 220 K. As
a result, a muonium atom that reaches the surface will be ejected
vertically \cite{TaqquPhysicsProcedia}. The small transverse momentum
it had before the ejection remains conserved or is reduced due to the
coupling to ripplon excitations so that the muonium atoms exit the
liquid with a very small divergency.

The experimental findings \cite{Muonium} are quite remarkable: The
outgoing velocity of the particles is 2180 m/sec with very small
dispersion which is in massive contradiction to the theoretical
result. Unless there are significant energy losses when the muonium
passes through the \he4 surface, one would conclude a chemical
potential of 32 K or 2.8 meV which is comparable to the hydrogen
chemical potential of 31 K or 2.6 meV.

Before we begin with the details of microscopic many-body
calculations, let us examine the two-body system consisting of a \he4
atom and \he3 or hydrogen-like atom. For the interaction between two
helium atoms we take the interaction of Aziz \etal from
Ref. \onlinecite{AzizIII}. For the He-H interaction we have examined
two models, the Lennard-Jones type interaction derived by Jochemsen
\etal\cite{Jochemsen84} and an earlier model by Toennies
\etal\cite{TWW76}.  The more recent work \cite{Jochemsen84} also
contains references to earlier interaction models. The parameters of
the interaction are determined by low-energy collisions \cite{TWW76}
and experimental high-energy scattering results
\cite{Jochemsen84}. The two interactions are numerically quite close,
they contain a strong repulsive core for an interparticle distance of
$\sigma = 3.19\,$\AA\ and a well depth of 0.566 meV \cite{TWW76} or
$\sigma = 3.2\,$\AA\ and a well depth of 0.6 meV \cite{Jochemsen84}.
Common to both of these interactions is the strongly repulsive core at
short distances. This is caused by both the Pauli repulsion acting
between the electrons, and their mutual Coulomb repulsion. The
analytic form of the interaction is in that region phenomenological.
We also include the same information for muonic and antiprotonic
helium which has received recently considerable attention
\cite{Antiprotons}.

To highlight information that is relevant for the properties of muonic
impurities in \he4 we first look at the zero-energy scattering
equation
  \begin{equation}
    -\frac{\hbar^2}{2\mu}\nabla^2\psi(r) + V^{\rm (IB)}(r)\psi(r) = 0
    \label{eq:scatteq}
  \end{equation}
of a \he3 and a hydrogen atom. $\mu = m_4 m_I/(m_4+m_I)$ is the
reduced mass, $m_4$ and $m_I$ are the \he4 and the impurity masses,
and $ V^{\rm (IB)}(r)$ is the interaction between a \he4 atom and the
\he3 or hydrogen-lile atom. The asymptotic wave function is
  \begin{equation}
    \psi(r) = 1-\frac{a_0}{r}\qquad{\rm as}\qquad r\rightarrow\infty
    \label{eq:scattpsi}
  \end{equation}
which defines the scattering length $a_0$. Another useful quantity for
comparing different interactions is the deBoer parameter \cite{deBoer}
  \begin{equation}
    \Lambda =
    \left(\frac{h^2}{\mu\epsilon\sigma^2}\right)^{1/2}\label{eq:deBoer}
      \end{equation}
where $\epsilon$ is the well depth and $\sigma$ is the core diameter.
Strictly, the deBoer parameter is defined specifically for 6-12
potentials, but the Aziz interaction is quite close to the 6-12
Lennard-Jones potential \cite{LennardJones35} and the Jochemsen
potential is close to the Toennies-Welz-Wolf interaction so that the
comparison remains meaningful.

Table \ref{tab:parameters} shows, for the \he3-\he4 system and the
six \he4-hydrogenic systems, the parameters of the calculated
scattering lengths $a_0$ from the Jochemsen interaction and from a
recent calculation \cite{HDTscatter} using the interaction by Meyer
and Frommhold \cite{MeyerFrommhold94}
\begin{table}[H]
\centerline{\begin{tabular}{c|cccccc}
    Atom & $\mu/m_4$& $\epsilon$ [meV] & $\sigma/a_B$ & $\Lambda$& $a_0/a_B$
    & $a_0/a_B$\cite{HDTscatter} \\
    \hline
    $^3$He    & 0.437 & 0.944 & 4.99 & 3.59 & -30.6 \\
    $\mu^4$He & 0.486 & 0.6   & 6.05 & 3.97 & 1050& \\
    $\bar {\mathrm p}^4$He & 0.50 & 0.6   & 6.05 & 3.71 &  66.6& \\
    T         & 0.437 & 0.6   & 6.05 & 4.07 & -29.9 & -37.6 \\
    D         & 0.341 & 0.6   & 6.05 & 4.61 & -7.93 & -9.09 \\
    H         & 0.205 & 0.6   & 6.05 & 5.93 & -0.17 & -0.228 \\
    Mu        & 0.028 & 0.6   & 6.05 & 15.98 & 3.11 & 
  \end{tabular}}
\caption{\label{tab:parameters} The table shows the reduced masses in
  units of the \he4 mass as well as the parameters of the interactions
  and the theoretical scattering lengths from the Jochemsen
  interaction \cite{Jochemsen84} and the Meyer-Frommhold interaction
  \cite{MeyerFrommhold94,HDTscatter} in units of the Bohr radius
  $a_B$.}
\end{table}
There is little experimental information on the scattering length of
Hydrogen and \he4. Chung and Dalgarno \cite{ChungDalgarno2002} cite a
value of $a_0 = 0.359\,a_B$ from the zero-temperature value of the
diffusion coefficients but also mention that direct calculation shows
it to be negative.

\begin{figure}[H]
\centerline{\includegraphics[width=0.4\textwidth,angle=270]{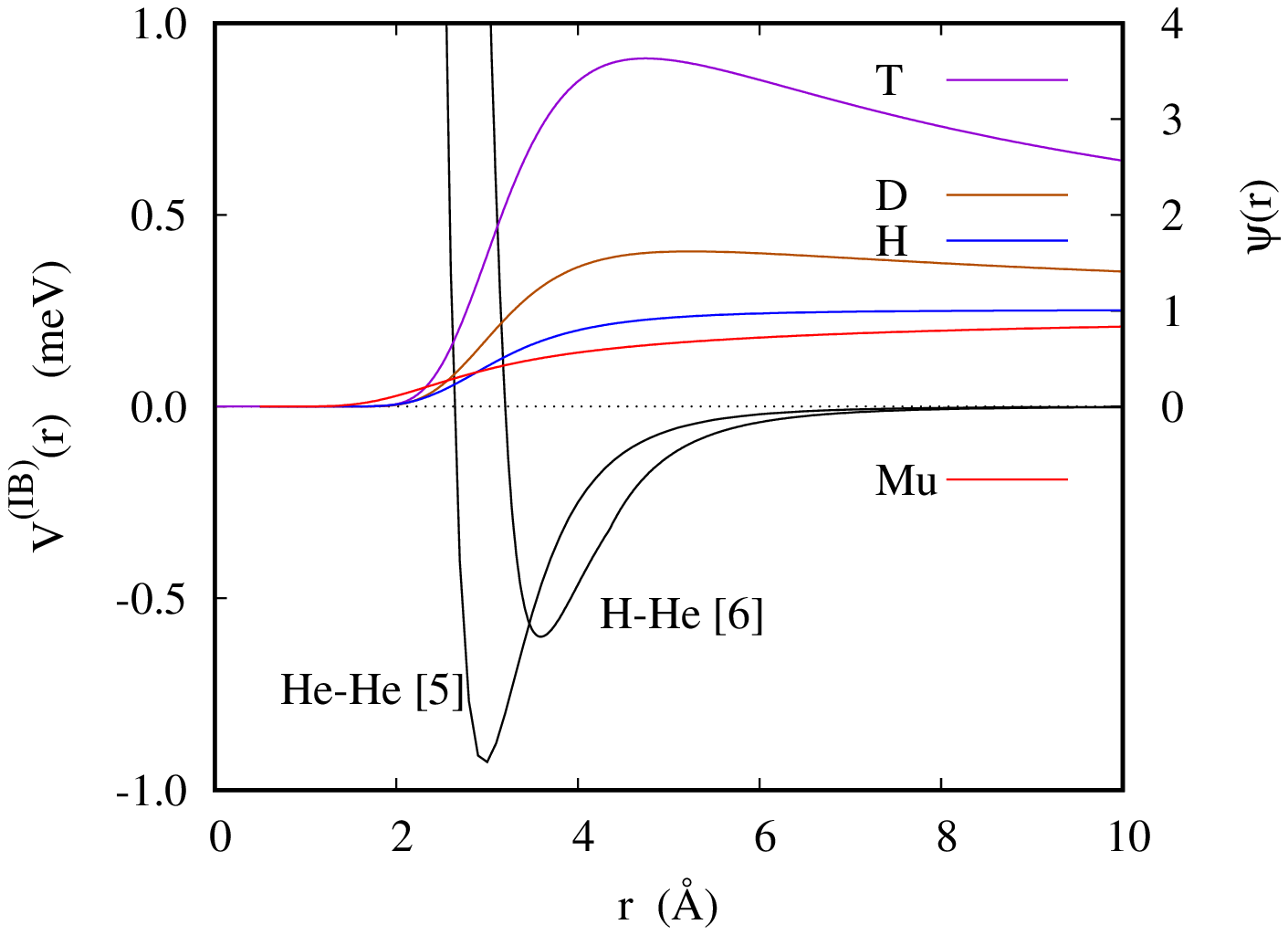}}
\caption{\label{fig:psiplot}(color online)
  The figure shows the Aziz interaction \cite{AzizIII} and the \he4-H
  interaction of Jochemsen \etal \cite{Jochemsen84} (left scale as indicated
  in the plot) as well as the s-wave scattering wave function $\psi(r)$
  (right scale).}
\end{figure}

Fig. \ref{fig:psiplot} shows the He-He interaction \cite{AzizIII}, the
He-H interaction \cite{Jochemsen84}, and the scattering wave functions
$\psi(r)$ for hydrogen, deuterium, tritium and muonium.  We do not
show the wave function for muonic and antiprotonic \he4, the above
interaction predicts a very weak bound state as indicated by the
positive scattering length. In fact muonic \he4 is close to a Feshbach
resonance.

Evidently the muonium atom is, in the sense of the deBoer parameter
$\Lambda$, by far the most ``quantum'' system examined here.

Attention is now directed towards the muonium wave function. Unlike
for other hydrogen isotopes, the wave function reaches, as a result of
the much larger zero-point motion of the muonium atom, very deep into
the repulsive core of the interaction which is poorly known. The
positive value of the scattering length stems from the overlap of the
wave function with that repulsive part of the interaction. We conclude
therefore that the large zero-point motion probes a rather
short-ranged part of the interaction. As a consequence, the results of
our many body calculations to be described below are affected by the
uncertainty of the interaction at these distances.

Our paper is organized as follows: In the next section
\ref{sec:structure} we outline very briefly the ideas of what we call
a ``generic'' many-body method. The method is called ``generic''
because, while we use here the Jastrow-Feenberg approach, the
equations can be derived in many different ways without even
mentioning a Jastrow wave function. For a recent review, see
Ref. \onlinecite{alphareexamined}.  We skip the technical details
which can be found in Ref. \onlinecite{EKthree} for the host liquid
and in Ref. \onlinecite{SKJLTP} for the impurity calculation.

Section \ref{sec:DMBT} then turns to the description of the dynamics
of a system that is governed by strong interactions such that mean-field
methods that are popular in condensed matter theory are inapplicable.
We spell out the guiding ideas and the final expression for
the dynamic density-density response function. The derivation
is found in Ref. \onlinecite{eomIII} and a comparison between theory
and experiment is Ref. \onlinecite{He4Dispersion}. The derivation
of the equations of motion for the impurity dynamics parallels very much
the one for the host liquid; details may be found in the supplemental
material \cite{eom4suppl}.

Our results will be discussed in section \ref{sec:results} and we close the
paper by a brief summary and outlook.

\section{Structure Methodology}
\label{sec:structure}

The full many-body Hamiltonian of $N$ $^4$He atoms with mass $m$ and
one impurity atom with mass $m_I$ is
\begin{eqnarray}
\label{eq:H}
  H_{N+1}^{(\rm I)}
  &=&  -\frac{\hbar^2}{2m_{\rm I}}\nabla_0^2 + \sum_{i=1}^{N}V^{\rm (IB)}(|\rvec_0-\rvec_i|) + H_N\, \label{eq:Himpu}\\
  H_N &=& 
  -\frac{\hbar^2}{2m}\sum_{i=1}^N\nabla_i^2
  + \sum_{\genfrac{}{}{0pt}{1}{i,j=1}{i<j}}^N V^{\rm (BB)}(|\rvec_i-\rvec_j|)\label{eq:Hhost}
  \end{eqnarray}
where $H_N$ is the Hamiltonian of the host system, and $m_{\rm I}$ and
$m_{\rm B}$ are the masses of the impurity atom and the background
(\he4) atom, respectively. As a convention, the impurity atom has the
coordinate $\rvec_0$ and the atoms of the host liquid the coordinates
$\rvec_1\cdots\rvec_N$.  To determine the ground state structure of
both the \he4 host liquid and the compound system, we use the
Jastrow-Feenberg {\em ansatz}\cite{FeenbergBook} for the many-body
wave function in conjunction with the Hypernetted chain/Euler-Lagrange
method to sum infinite classes of diagrams and to optimize the
correlations. The method is a well established, fast, and accurate
method for calculating properties of strongly interacting quantum
fluids, it has been described in numerous review articles and
pedagogical material, see, for example,
Ref.~\onlinecite{MikkoQFSBook}.

\subsection{\he4 background methodology}
\label{ssec:HostEnergy}
The Jastrow-Feenberg form of the wave of an $N$-body system is in the
limit $N\rightarrow\infty$ written in terms of multi-particle
correlation functions $u_n(\rvec_1,\ldots,\rvec_n)$; truncation at
$n=3$ is normally sufficient.
\begin{widetext}
\begin{equation}
        \Psi_N({\bf r}_1,\ldots,{\bf r}_N) =
        \exp\frac{1}{2}\Bigl[
         \sum_{1\le i<j\le N} u_2({\bf r}_i,{\bf r}_j)
        + \sum_{1\le i<j<k\le N} u_3({\bf r}_i,{\bf r}_j,{\bf r}_k)
        + \ldots\Bigr].
\label{eq:Jastrow}
\end{equation}
\end{widetext}

The correlation functions $u_n({\bf r}_1,\ldots,{\bf r_n})$ are
determined by minimization of the $N$-body energy-expectation value
\begin{equation}
        E_N = \frac{\langle\Psi_N\vert H_N \vert \Psi_N\rangle}{
        \langle\Psi_N\vert\Psi_N\rangle}.
\label{eq:EN}
\end{equation}
through
\begin{equation}
        \frac{\delta E_N}{\delta u_n({\bf r}_1,\ldots,{\bf r_n})} = 0.
\label{eq:eulerHost}
\end{equation}
The energy expectation value and other physically relevant quantities
are calculated by diagrammatic expansions and summations of infinite
classes of diagrams. The hierarchy of ``hypernetted chain'' integral
equations \cite{Morita58,LGB} provides a scheme that is, at every
level of implementation, consistent with the optimization problem
(\ref{eq:eulerHost}) in the sense that the resulting pair distribution
and structure functions reproduce {\it qualitatively\/} the properties
of the solution of the exact variational problem.  It should be noted
that the procedure is equivalent to a self-consistent summation of all
ring-- and ladder--diagrams in Feynman's perturbation theory
\cite{parquet1,parquet2,parquet3}, the so-called parquet diagrams.

For comparison with the impurity calculations in the next section
\ref{ssec:ImpuEnergy} and further discussions, we need to spell out
the energy expression and Euler equation resulting from the variational
problem \eqref{eq:eulerHost}. The theory is formulated in terms of the
physically observable pair distribution function
\begin{eqnarray}
g_{\rm BB}(\rvec,\rvec')&=&g_{\rm BB}(|\rvec-\rvec'|)\label{eq:gBB}
\\
&\equiv&\frac{1}{\rho^2}
\frac{\bra{\Psi_N}\sum_{i\ne j}\delta(\rvec_i-\rvec)\delta(\rvec_j-\rvec')
\ket{\Psi_N}}{\ovlp{\Psi_N}{\Psi_N}}\nonumber
\end{eqnarray}
and the static structure function
\begin{equation}
 S_{\rm BB}(k) = 1 + \rho\int d^3r e^{\I{\bf k}\cdot{\bf
     r}}\left[g_{\rm BB}(r)-1\right]
\label{eq:sBB}
\end{equation}
where $\rho$ is the particle density of the system. For further
reference, we spell out the explicit form:
\begin{equation}
E = E_{\rm R}+ E_{\rm Q} + E_{\rm ele} ,
\label{eq:etotHost}\end{equation}
where
\begin{eqnarray}
\frac{E_{\rm R}}{N} &=& \frac{\rho}{2}\int d^3r \left[v(r) g_{\rm BB}(r)
+ \frac{\hbar^2}{m}\left|\nabla \sqrt{g_{\rm BB}(r)}\right|^2\right],\nonumber\\
\label{eq:eRHost}\\
E_{\rm Q} &=& -\frac{N}{4} \int \frac{d^3 k}{ (2\pi)^3\rho} t(k)
\frac{(S_{\rm BB}(k)-1)^3}{S_{\rm BB}(k)},
\label{eq:eQHost}
\end{eqnarray}
in which $t(k) = \hbar^2 k^2/2m$, and $E_{\rm ele}$ is the contribution
from irreducible diagrams and higher correlation functions
$u_n({\bf r}_{1},.., {\bf r}_{n})$ for $n\ge 3$ which can be expressed
entirely in terms of the  the pair distribution function $g_{\rm BB}(r)$
Its detailed structure \cite{EKthree} is irrelevant for the remaining
discussion.

The Euler equation for $g_{\rm BB}(r)$ can be formulated in different
ways \cite{FeenbergBook,ChuckReview,EKthree,LanttoSiemens},
for the sake of discussion we write it in coordinate space \cite{LanttoSiemens}
as
\begin{eqnarray}
  &&\frac{\hbar^2}{ m}\nabla^2\sqrt{g_{\rm BB}(r)}\label{eq:bbgHost}
\\
  &=&\left[V^{\rm (BB)}(r)
  + V^{\rm (BB)}_{\mathrm{ele}}(r) + w^{\rm (BB)}_{\mathrm
    I}(r)\right]\sqrt{g_{\rm BB}(r)},\nonumber
\end{eqnarray}
where $V^{\rm (BB)}_{\mathrm{ele}}(r)$ originates from  $E_{\rm AA}$, and
\begin{equation}
{\tilde w}_{\mathrm I}(k) = - \frac{t(k)}{2}\left[\frac{1}{ S_{\rm BB}^2(k)}-1\right]
-t(k)\left[S_{\rm BB}(k)-1\right].
\label{eq:windHost}
\end{equation}
As usual in this field, we have defined the dimensionless Fourier transform
by including a particle number density factor $\rho$:
\begin{equation}
{\tilde f(k)} \equiv \rho\int d^3r\, e^{\I \kvec\cdot\rvec}f(r)\label{eq:ft}.
\end{equation}
The important point to notice here is that the ``induced interaction''
is, no matter what  $V^{\rm (BB)}_{\mathrm {ele}}(r)$ is, just right so that the
resulting $\sqrt{g_{\rm BB}(r)}$ has zero scattering length, \ie
that \cite{FeenbergBook}

\begin{equation}
  g_{\rm BB}(r) = 1 + {\cal O}(r^{-4})\qquad {\rm  as\qquad} r\rightarrow\infty
  \label{eq:gBBlong}\,.
\end{equation}

For long wavelengths, the static structure function goes as
\begin{equation}
  S_{\rm BB}(k) = \frac{\hbar k}{2mc_s}\quad\mbox{as}\quad k\rightarrow 0
\label{eq:Slong}
\end{equation}
where $c_s$ is the speed of sound that is obtained from a theory of excitations
or from the equation of state
\begin{equation}
mc_s^2 = \frac{d}{d\rho}\rho^2\frac{d (E_N/N)}{d\rho}\,.
\label{eq:mceos}
\end{equation}
The $c_s$ obtained from the equation of state through the density
derivative \eqref{eq:mceos} will normally be {\it different\/} from
the speed of sound obtained from the slope of $S_{\rm BB}(q)$ from
Eq. \eqref{eq:Slong}. In fact these two quantities agree only in an
exact theory \cite{EKVar,parquet5,PhysRevA.110.052222}. The
inconsistency can be used as a measure for the convergence of our
diagrammatic expansions.

\subsection{Singe impurity methodology}
\label{ssec:ImpuEnergy}

In the next step, the impurity is included. The wave function of the
compound system is
\begin{widetext}
\begin{equation}
        \Psi_N^{(I)} (\rvec_0,\rvec_1,...,\rvec_N)
        = \exp \frac{1}{2} \Bigl[ \sum_{j=1}^N u^{I}(\rvec_0,\rvec_j)
        +\sum_{1\le j<k\le N}
        u^{I}(\rvec_0,\rvec_j,\rvec_k)
        \Bigr]\Psi_N(\rvec_1,...,\rvec_N).
\nonumber\\
\label{eq:waveI}
\end{equation}
\end{widetext}
The {\it chemical potential\/} of the impurity is the energy gained or
lost by adding one impurity particle into the liquid, in other words it
equals to the energy difference
\begin{eqnarray}
        \mu^{(I)} &=& E_{N+1} - E_N
\nonumber\\
&=& \frac{\langle\Psi^{(I)}_N\vert H^I_{N+1} \vert \Psi^{(I)}_N\rangle
                }{ \langle\Psi^{(I)}_N\vert\Psi^{(I)}_N\rangle}
        -\frac{\langle\Psi_N\vert H_N \vert \Psi_N\rangle}{
        \langle\Psi_N\vert\Psi_N\rangle}.
\label{eq:chemi}
\end{eqnarray}
The Euler equation is then derived by minimizing the impurity chemical
potential
\begin{equation}
        \frac{\delta \mu^{(I)}}{\delta u_n^{I}({\bf r}_0,\ldots,{\bf r_n})} = 0.
\label{eq:eulerImpu}
\end{equation}

Results are again the energetics as well as structure and
distribution functions \cite{SKJLTP}. The impurity-background
distribution function is
\begin{eqnarray}
&&g_{\rm IB}(\rvec,\rvec')=g_{\rm IB}(|\rvec-\rvec'|)\label{eq:gIB}
\\
&\equiv&\frac{1}{\rho}
\frac{\bra{\Psi_N^{(I)}}\sum_{1\le1 \le N}\delta(\rvec_0-\rvec)\delta(\rvec_i-\rvec')
\ket{\Psi_N^{(I)}}}{\ovlp{\Psi_N^{(I)}}{\Psi_N^{(I)}}}\nonumber
\end{eqnarray}
and the static impurity-background structure function
\begin{equation}
 S_{\rm IB}(k) = \rho\int d^3r e^{\I{\bf k}\cdot{\bf
     r}}\left[g_{\rm IB}(r)-1\right]
\label{eq:sIB}
\end{equation}

The chemical potential
$\mu^{(I)}$ takes a form similar to Eq. \eqref{eq:etotHost},
  \eqref{eq:eRHost}, \eqref{eq:eQHost}, 
\begin{eqnarray}
  \mu^{(I)}&=& \mu_R^{(I)}+\mu_Q^{(I)}+\mu_{\rm ele}^{(I)}\label{eq:muImpu}\\
   \mu_R^{(I)} &=& \rho \int d^3r
\Biggl[ g_{\rm IB}(\rvec)
V^{\rm (IB)}(r) + \frac{\hbar^2}{\mu}
\left|\nabla\sqrt{ g_{\rm IB}(r)}\right|^2\Biggr]\nonumber\\
  &\equiv& \mu_V^{(I)} + \mu_T^{(I)}\label{eq:eRImpu}\\
  \mu_Q^{(I)}&=&
\frac{1}{2}\int \frac{d^3k}{(2\pi)^3\rho}
S_{\rm IB}(k) \tilde w^{\rm (IB)}_{\rm I}(k)
\label{eq:eQImpu}
\end{eqnarray}
where
\begin{equation}
  \tilde w^{\rm (IB)}_{\rm I}(k) = -
         \frac{S_{\rm IB}(k)(S_{\rm BB}(k)-1)}{4S_{\rm BB}(k)}
\left(\frac{\hbar^2 k^2}{\mu} + \frac{\hbar^2 k^2}{m S_{\rm BB}(k)}
\right)\,.
\label{eq:windImpu}
\end{equation}
is the induced potential.
$\mu_{\rm ele}^{(I)}$
is again the  correction from elementary diagrams and multiparticle
correlations; its calculation is tedious \cite{SKJLTP} but the specific form
is of no relevance for the further discussion. The Euler equation
for the impurity-background distribution function is analogous to
Eq. \eqref{eq:bbgHost}
\begin{eqnarray}
&&\frac{\hbar^2}{\mu}\nabla^2
  \sqrt{g_{\rm IB}(r)}\nonumber\\
  &=& \left[V^{\rm (IB)}(r) + w^{\rm (IB)}_{\rm I}(r)
+ V^{\rm (IB)}_{\rm ele}(r)
 \right]
\sqrt{g_{\rm IB}(r)}\,.\label{eq:bbgImpu}
\end{eqnarray}
where $V^{\rm (IB)}_{\rm ele}(r)$ originates again from elementary
diagrams and triplet correlations.  The effect of the induced
interaction $w^{\rm (IB)}_{\rm I}(r)$ is, among others, again that the
$\sqrt{ g_{\rm IB}(r)}$ has zero scattering length.

There is again a hydrodynamic consistency condition between the long
wavelength limit of the static structure function and the density
dependence of the chemical potential: The long wavelength limit of the
impurity--background structure function is
\cite{FeenbergBook}
\begin{equation}
S_{\rm IB}(k) = -\alpha \quad\mbox{as}\quad k\rightarrow 0
\label{eq:alphaSofq}
\end{equation}
where $\alpha$ is the volume excess factor
\begin{equation}
\alpha = \frac{\rho}{mc_s^2}\frac{d \mu^{I}}{d\rho}
\label{eq:alphaEOS}
\end{equation}
which represents the relative increase in volume when a \he4 atom is
replaced by an impurity.  The volume excess factor $\alpha$ can be
obtained directly from $S_{\rm IB}(0+)$ or from the density dependence
of the impurity chemical potential. Again, the long wavelength limit
(\ref{eq:alphaSofq}) and the hydrodynamic derivative
(\ref{eq:alphaEOS}) agree only in an exact theory.  From the point of
view of diagrammatic expansions of both the chemical potential
$\mu^{(I)}$ and the particle-hole interaction it can be shown that the
derivatives $\frac{d \mu^{I}}{d\rho}$ and \eqref{eq:mceos} contain a
{\em proper superset\/} of the diagrams contributing to the limit
\eqref{eq:Slong}\cite{EKVar}; hence one should generally consider the
expression \eqref{eq:alphaEOS} to be provide a more accurate
expression for the volume excess factor.

\section{Many-Body Dynamics}
\label{sec:DMBT}

\subsection{Dynmics of the host liquid}
\label{ssec:HostDMBT}

Dynamics is treated at the same level as the ground state:
The system is exposed to a small, local and time-dependent external field
\begin{equation}
  H_{\rm ext} = \sum_{1\le i\le N}V_{\rm ext}(\rvec_i,t)
    \label{eq:HextHost}\,.
\end{equation}
The dynamic wave function acquires a small, time-dependent
component which we write in the form
\begin{eqnarray}
        \left|\Psi(t)\right\rangle&=&e^{-\I E_Nt/\hbar}
          \frac{e^{\half\delta U(t)}
        \left|\Psi_N\right\rangle}{[\langle\Psi_N|e^{\half\delta U^\dagger(t)}
              e^{\half\delta U(t)}|\Psi_N\rangle]^{1/2}}\nonumber\\
          &\equiv& e^{-\I E_Nt/\hbar}
          \left|\Psi_0(t)\right\rangle \ ,
\label{eq:DynHost}
\end{eqnarray}
where $\delta U(t)$ is an excitation operator that is written, for the
case of bosons, in the same manner as the ground state wave function:
\begin{eqnarray}
        \delta U(t)
        &=&\sum_{1\le i\le N}\delta u_1(\rvec_i;t)
        +\sum_{1\le i<j\le N}
        \delta u_2(\rvec_i,\rvec_j;t)\nonumber\\
        &+&\sum_{1\le i<j<k\le N}\delta u_3({\bf r}_i,{\bf r}_j,{\bf r}_k;t)+
        \ldots\,.
\label{eq:deltaUHost}
\end{eqnarray}
The amplitudes $\delta u_i(\rvec_1,\ldots\rvec_i;t)$ are determined
by the time-dependent generalization of the Ritz' variational principle:
\begin{equation}
  \frac{\delta}{\delta u_i(\rvec_1,\dots\rvec_i;t)} \int\!\!
  dt\,\langle\,\Psi(t)|H_N + H_{\rm ext}
    -\I\hbar\partial_t|\Psi(t)\rangle = 0\,.
\label{eq:eomHost}
\end{equation}

Assuming that $\delta U(t)$ is a small perturbation of the ground
state correlations, we can linearize the equations of motion for
$\delta u_i(\rvec_i,\dots;t)$, leading to the density--density
response function $\chi(k,\omega)$ from which we obtain the dynamic
structure function $S(k,\omega)=\Im \,\chi(k,\omega)$:
\begin{eqnarray}
\chi(k,\omega) &=&
        \frac{S_{\rm BB}(k)}{\hbar \omega - \varepsilon_{\rm F}(k) - \Sigma^{\rm (B)}(k,\hbar\omega)}
        \nonumber\\&+&
        \frac{S_{\rm BB}(k)}{-\hbar \omega -\varepsilon_{\rm F}(k) - \Sigma^{\rm (B)}(k,-\hbar\omega)}\ ,
\label{eq:chiHost}
\end{eqnarray}
where $\varepsilon_{\rm F}(k) = \hbar^2 k^2/2m S_{\rm BB}(k)$ is the
Feynman excitation spectrum, and the self-energy is given by an
integral equation
\begin{widetext}
\begin{equation}
        \Sigma^{(\rm B)}(k,\hbar\omega) =
        \frac{1}{2}
        \int \frac{d^3p_1d^3p_2}{(2\pi)^3\rho}\,
        \frac{\delta({\bf k}-{\bf p}_1-{\bf p}_2)
        \left|\tilde V_3({\bf k};{\bf p}_1,{\bf p}_2)\right|^2}
        {\hbar\omega-\varepsilon_{\rm F}(p_1)
        -\Sigma^{(\rm B)}(p_1,\hbar\omega-\varepsilon_{\rm F}(p_2))
        -\varepsilon_{\rm F}(p_2)-\Sigma^{(\rm B)}(p_2,\hbar\omega-\varepsilon_{\rm F}(p_1))}\,.
\label{eq:sigmaHost}
\end{equation}
$\tilde V_3(\kvec;\pvec_1,\pvec_2)$ is the three-phonon vertex
\begin{equation}
       \tilde V_3({\bf k};{\bf p}_1,{\bf p}_2)
       =\frac{\hbar^2}{2m}\sqrt{\frac{S_{\rm BB}(p_1)S_{\rm BB}(p_2)}{S_{\rm BB}(k)}}
        \left[
        {\bf k}\cdot {\bf p}_1\tilde X_{\rm BB}(p_1)+{\bf k}\cdot {\bf p_2}\tilde X_{\rm BB}(p_2)
        - q^2\tilde X_{\rm BBB}({\bf q},{\bf k}_1,{\bf k}_2)\right]\,,
\label{eq:V3}
\end{equation}
\end{widetext}
where $\tilde X_{\rm BB}(k) = 1-1/S_{\rm BB}(k)$ is the ``direct correlation function'', and
$\tilde X_{\rm BBB}({\bf k},{\bf p}_1,{\bf
  p}_2)$ is the fully irreducible three-phonon coupling matrix
element. In the simplest approximation, $\tilde X_{\rm BBB}({\bf q},{\bf
  k}_1,{\bf k}_2)$ is replaced by the three--body correlation $\tilde
u_3(\kvec,\pvec_1,\pvec_2)$; this approximation ensures that
long--wavelength properties of the excitation spectrum are preserved
\cite{Chuckphonon}. The CBF-Brillouin-Wigner (CBF-BW)
approximation \cite{JacksonSkw} is obtained by omitting the
self-energy corrections in the energy denominator of
Eq. (\ref{eq:sigmaHost}).

The implementation of the method outlined only briefly here has led to
an unprecedented agreement between theoretical predictions \cite{eomIII}
and experimental results \cite{skw4lett,skwpress}.

\subsection{Dynamics of the impurity motion}
\label{ssec:ImpuDMBT}

In analogy to section \ref{ssec:HostDMBT} the system is exposed to
a small external field $H_{\rm ext}^{(I)}$ which couples to the impurity
only,
\begin{equation}
  H_{\rm ext}^{(I)} = V_{\rm ext}(\rvec_0,t)
\end{equation}
The time dependent wave function is, in analogy to Eq. \eqref{eq:DynHost}
\begin{eqnarray}
        \ket{\Psi^{(I)} (t)} &=& e^{-\I E_{N+1}^I t / \hbar }
        \frac {e^{\half \delta U(t)}\ket{\Psi_N^{(I)}}}
            {[\langle\Psi_N^{(I)}|e^{\half\delta U^\dagger(t)}
                e^{\half\delta U(t)}|\Psi_N^{(I)}\rangle]^{\half}}\nonumber\\
            &\equiv&  e^{-iE_{N+1}^I t / \hbar } \ket{\Psi_0^{(I)} (t)}\,,
            \label{eq:DynImpu}
\end{eqnarray}
\begin{eqnarray}
        \delta U(t)
        &=&
        \delta u_1({\bf r}_0;t) +  \sum_{1\le i\le N}
        \delta u_2({\bf r}_0,{\bf r}_i;t)\nonumber\\& +&
	\sum_{1\le i<j\le N}\delta u_3({\bf r}_0,{\bf r}_i,{\bf r}_j;t)+\ldots
\label{eq:deltaUImpu}
\end{eqnarray}
and determined by the stationarity principle
\begin{equation}
  \frac{\delta}{\delta u_i(\rvec_0,\dots\rvec_i;t)} \int\!\!
  dt\,\langle\,\Psi(t)|H_{N+1}^I + H_{\rm ext}^{(I)}
    -\I\hbar\partial_t|\Psi(t)\rangle = 0\,.
\label{eq:eomimpu}
\end{equation}
The derivation of a workable form of the equations of motion is very
similar to the one for the host system, details will carried out in
the supplemental material to this work \cite{eom4suppl}. The impurity
dispersion relation is, as usual, determined by a self-energy
\begin{equation}
  \hbar\omega(k) = \varepsilon_I(k)+\Re\Sigma^{(I)}(k,\omega(k))
\label{eq:Dispersion}
\end{equation}
where $\varepsilon_I(q) = \hbar^2 q^2/2m_I$ is the kinetic
energy of the free impurity particle and $\Sigma^{(I)}(q,\omega)$ its self-energy.

We will derive 2 versions, an ``unrenormalized'' and a ``renormalized'' form.
The essential building block is the unrenormalized self-energy
\begin{widetext}
\begin{equation}
  \Sigma_u^{(I)}(k,\omega)=\int \frac{d^3p}{(2\pi)^3\rho}
\frac{S_{\rm BB}(p)\left|\displaystyle\hm2I\kvec\cdot\pvec\tilde X_{\rm IB}(p)\right|^2}
     {\hbar\omega - \varepsilon_I(\kvec+\pvec)-\Sigma^{(I)}(|\kvec+\pvec|,\hbar\omega-\varepsilon_F(k)) - \varepsilon_F(p)-\Sigma^{(B)}(p,\hbar\omega-\varepsilon_I(p))}\label{eq:sigmau}
\end{equation}
\end{widetext}
where $\tilde X_{\rm IB}(p)= (S_{\rm IB}(p)-1)/S_{\rm BB}$.  The
unrenormalized form identifies
$\Sigma^{(I)}(k,\omega)\equiv\Sigma_u^{(I)}(k,\omega)$ whereas the
renormalized form, which sums a larger class of diagrams, identifies
\begin{equation}
  \Sigma^{(I)}(k,\omega) \equiv \Sigma_r^{(I)}(k,\omega) = \frac{\Sigma_u^{(I)}(k,\omega)}{1-\displaystyle\frac{1}{\varepsilon_I(k)}
    \Sigma_u^{(I)}(k,\omega)}\label{eq:sigmar}\,.
\end{equation}
If we omit three-body and higher order fluctuations in the excitation
operator \eqref{eq:deltaUImpu}, the self-energy reduces to
Eq. (3.29) of Ref. \onlinecite{SKJLTP},

\begin{equation}
  \Sigma_u(k,\omega) = \int \frac{d^3p}{(2\pi)^3\rho}S_{\rm BB}(p)
  \frac{\left[\displaystyle\frac{\hbar^2}{2m_I}\kvec\cdot\pvec
      \tilde X_{\rm IB}(p)\right]^2}{\hbar\omega - \varepsilon_I(|\kvec+\pvec|)-\varepsilon_B(p)}\,.\label{eq:sigmau2}
    \end{equation}
The form \eqref{eq:sigmau} is a quite plausible generalization of
Eq. \eqref{eq:sigmau2}, the only non-trivial aspect is at which energy
and momentum the self-energy corrections in the denominator of
Eq. \eqref{eq:sigmau} are to be taken. The modification of the energy
denominator, is, of course, important: If we omit multiparticle
fluctuations, the energy denominator of Eq. \eqref{eq:sigmau2}
describes the coupling of a non-interacting impurity to a Feynman
phonon, whereas the energy denominator of Eq. \eqref{eq:sigmau} the
coupling of the physical impurity to the physical phonon-roton
spectrum.

Expressions \eqref{eq:sigmau} and \eqref{eq:sigmau2} are, of course,
reminiscent of the G(0)W approximation \cite{Hedin65,Rice65} but they
deviate in the important aspect that the self-energy goes to zero in
the zero momentum limit which is a consequence of the  fact that we
have built the excitation theory on an interacting ground state. We
will get back to this point further below when we discuss the meaning
of the renormalized vs. unrenormalized form of the
self-energy.

In the long wavelength limit, we get two explicit $\qvec$ factors from
the self-energy \eqref{eq:sigmau} and the self-energy becomes
independent of the energy $\hbar\omega$. Thus we get
\begin{equation}
  \Sigma_u^{(I)}(k,\omega) = -\varepsilon_I(k) I
 \label{eq:sigmau0}
\end{equation}
with
\begin{widetext}
\begin{equation}
  I 
=\frac{1}{3}\int\frac{d^3p}{3(2\pi)^3\rho}\frac{S_{\rm BB}(p)\varepsilon_I(p) \tilde X_{\rm IB}^2(p)}
     {\varepsilon_I(p)+\Sigma^{(I)}(p,0) + \varepsilon_B(p)+\Sigma^{(B)}(p,-\varepsilon_I(p))}\,.\label{eq:Idef}
\end{equation}
\end{widetext}
Evidently, the correction involves only self-energies for negative energy
which guarantees that the spectrum is real in the long
wavelength limit. The long wavelength limit \eqref{eq:sigmau0}
predicts a spectrum of the expected form
\begin{equation}
  \hbar\omega(k) = \frac{\hbar^2 k^2}{2m^*_I}\qquad\mathrm{as}\qquad k\rightarrow 0+\label{eq:epslong}
  \end{equation}
where the effective mass is
\begin{equation}
\frac{m^*_I}{m_I}\Biggr)_{u} = \frac{ 1 }{ 1 - I}
\label{eq:munren}
\end{equation}
or
\begin{equation}
\frac{m^*_I}{ m_I}\Biggr)_{r} =  1 + I
\label{eq:muren}
\end{equation}
are the effective mass predicted by the unrenormalized and
the renormalized form of the self-energy,
respectively. In the case \eqref{eq:sigmau2}
we obtain
\begin{equation}
  I \rightarrow I_0 =
  \frac{1}{3}\int\frac{d^3p}{(2\pi)^3\rho}\frac{S_{\rm
      BB}(p)\varepsilon_I(p) \tilde X_{\rm IB}^2(p)} {\varepsilon_I(p)
    + \varepsilon_B(p)}\,.
\end{equation}
Eqs. \eqref{eq:muren} and \eqref{eq:munren}
  have, for that case, first been derived by Owen
  \cite{OWE81} who also coined the names ``unrenormalized'' and
  ``renormaized''.

We conclude this section by briefly discussing the relationship to the
G(0)W approximation \cite{Hedin65,Rice65} which is in condensed matter
theory a popular procedure to discuss the self-energy and the
effective mass (\ref{eq:munren}). We start from a Green's functions
expression for the impurity self-energy:
\begin{eqnarray}
  &&\Sigma_{\rm GW}(\kvec,E) \label{eq:greenself}
  \\
  &=&\I\int \frac{d^3 p d(\hbar\omega)}{(2\pi)^4\rho^B}
	G(\kvec-\pvec,E-\omega) \left[\tilde V_{\rm p-h}^{\rm (IB)}(p)\right]^2
	\chi(\pvec,\omega)
\nonumber
\end{eqnarray}
where $G(\kvec,E)$ is the Green's function of a single impurity
particle, $\tilde V_{\rm p-h}^{\rm (IB)}(\pvec)$ the coupling matrix element
between the impurity and a phonon, and $\chi(\pvec,\omega)$
the density-density response function of the background.
Of course, the theory at this level cannot determine
the coupling matrix element $\tilde V_{\rm p-h}^{(IB)}(q)$.

To make the connection with the above derivations, we assume for
simplicity that the excitations of the background are exhausted by the
Feynman phonon, \ie we have
\begin{equation}
	\chi(\kvec,\omega) = \frac{\hbar^2 k^2/m }{
	\hbar^2\omega^2 - \varepsilon_B^2(k) + \I \eta }.
\end{equation}
Then, we can carry out the frequency integration and find
\begin{eqnarray}
  &&\Sigma_{\rm GW}(\kvec,\omega)\label{eq:sigmaGW}\\
  &=&\int \frac{d^3 p}{(2\pi)^3\rho}
	G_0(\kvec-\pvec,\hbar\omega-\varepsilon_B(p))
	\left[V_{\rm p-h}^{\rm (IB)}(p)\right]^2 S_{\rm BB}(p).
\nonumber
\end{eqnarray}
Moreover, we replace the full single-particle Green's function
in Eq. (\ref{eq:greenself}) by the free particle Green's function
\begin{equation}
	G_0(\kvec,\omega) = \frac{1}{\hbar\omega - \varepsilon_I(\kvec) + \I\eta}.
\label{eq:green0}
\end{equation}
and use the relationship \cite{SKJLTP}
\begin{equation}
S_{\rm IB}(k) = -2\frac{V^{\rm (IB)}_{\rm p-h}(k)S_{\rm BB}(k)}{
        \varepsilon_I(k) + \varepsilon_B(k)},
\label{eq:IVPH}
\end{equation}
between the impurity-background structure function $S_{\rm IB}(k)$
and the impurity-phonon coupling matrix element $\tilde V_{\rm p-h}^{(IB)}(p)$.
The self-energy $\Sigma_{\rm GW}(q,\omega)$ has, for small $q$,
the quadratic form
\begin{equation}
  \Sigma_{\rm GW}(k,\omega) \rightarrow \Sigma_{\rm GW}(0,0)
  + \varepsilon_I(k) I_0\qquad{\rm for}\qquad (k,\omega)\rightarrow 0\,.
\label{eq:sigmaGWons}
\end{equation}
\ie  we recover the
unrenormalized effective mass (\ref{eq:munren}).

If we, on the other hand, use $E = \varepsilon^*_I(k) = \frac{\hbar^2
  k^2}{2m_I^*}$ in the Green's function \eqref{eq:green0}
we obtain the renormalized form
\begin{equation}
\frac{m^*_I}{ m_I}\Biggr)_{r} =  1 + I_0^*
\end{equation}
where
\begin{equation}
  I_0^* =
  \frac{1}{3}\int\frac{d^3p}{(2\pi)^3\rho}\frac{S_{\rm BB}(p)\varepsilon_I(p) \tilde X_{\rm IB}^2(p)} {\varepsilon_I^*(p)
    + \varepsilon_B(p)}\,.\label{eq:Istar}
    \end{equation}
In this case, $m^*$ would have to be determined self-consistently.
But of course, the assumption of a constant effective mass for
all momentum transfers where the integrand of Eq. \eqref{eq:Istar}
is sizable is rather crude.

\section{Results}
\label{sec:results}

\subsection{Structure}
\label{ssec:StaticResults}
In our numerical applications we assume that the \he4 atoms interact
via the Aziz potential \cite{AzizII}. This potential reproduces the
equation of state of \he4 with high accuracy \cite{JordiEncyclopedia}. For
the interaction $V^{\rm (IB)}(r)$ between the host liquid and the
hydrogenic impurity atoms we have used the potential derived by
Jochemsen {\em et al.\/} \cite{Jochemsen84}; results obtained with the
interaction derived earlier by Toennies {\em et al.\/}\cite{TWW76} for
the hydrogen isotopes are almost indistinguishable \cite{SKJLTP}.
\begin{figure*}[t]
  \centerline{\includegraphics[width=0.5\textwidth,angle=270]{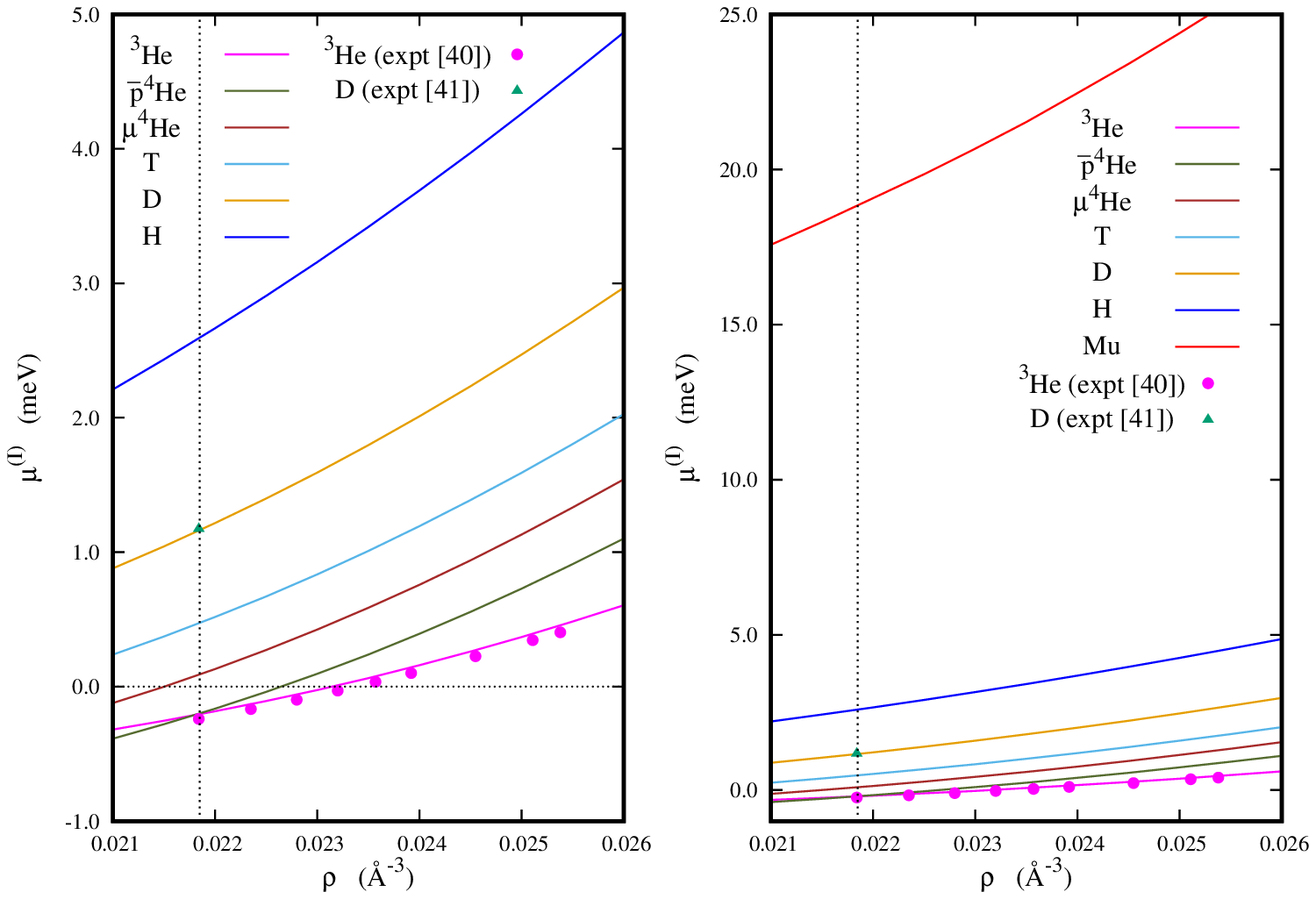}}
    \caption{(color online) The left figure shows the chemical
      potential of \he3 impurities, muonic \he4, antiprotonic \he4 as
      well as hydrogen isotopes in \he4. Experimental data are from
      Refs. \onlinecite{EbnE} and \onlinecite{Reynolds91}, the
      vertical dotted line indicates the equilibrium density of \he4,
      $\rho_0 = 0.02185\,$\AA$^{-3}$.
      The horizontal dotted line at zero chemical potential is a guide
      to the eye. The right figure shows, on a different scale, the
      chemical potentials of the same H isotopes as well as the
      chemical potential of muonium.\label{fig:eosplot}}
\end{figure*}

The first result of our calculations is, of course, the impurity
chemical potential. Figs. \ref{fig:eosplot} shows our results together
with available experimental data for \he3 \cite{EbnE} and D
\cite{Reynolds91}; the results for \he3 and the hydrogen isotopes are
identical to those of Ref. \onlinecite{SKJLTP}. The
chemical potential of muonium is almost an order of magnitude larger
than that of all other impurity atoms which is due to the much larger
zero-point motion.

\begin{figure}[H]
  \centerline{\includegraphics[width=0.3\textwidth,angle=270]{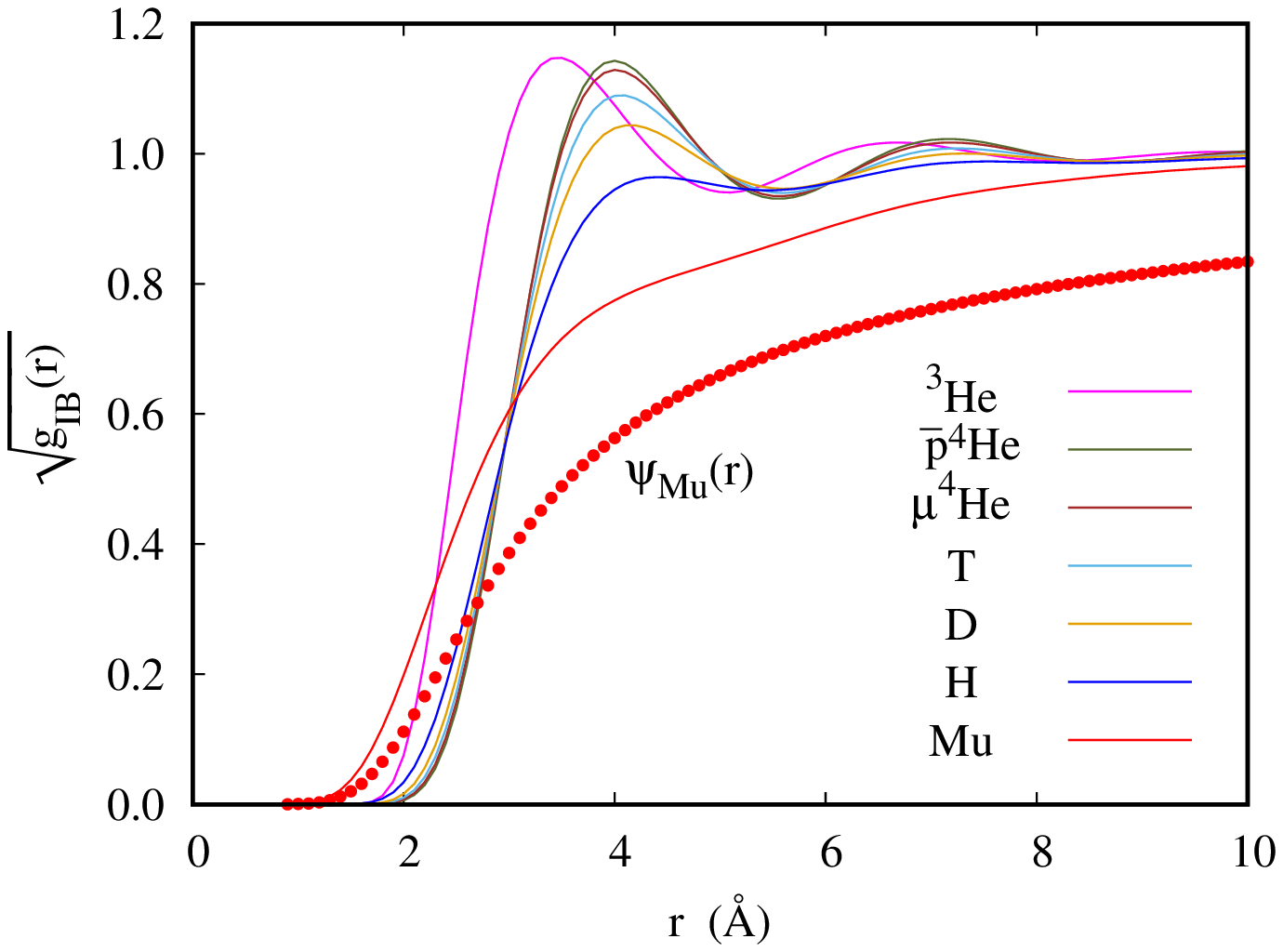}}
  \caption{(color online) The figure shows $\sqrt{g_{\rm IB}(r)}$ for
    \he3, hydrogen isotopes, muonic helium, antiprotonic helium, and
    muonium. The dotted line shows, for comparison, the pair wave
    function $\psi_{\rm Mu}(r)$ of Fig. \ref{fig:psiplot} for
    muonium. \label{fig:gibplot}}
\end{figure}

While the reliability of our methods for \he3 and hydrogen impurities
is well tested, the large zero-point motion of muonium raises of
course some concerns. To demonstrate the issue, we show in
Fig. \ref{fig:gibplot} the square-root of the impurity-background
distribution function $\sqrt{g_{\rm IB}(r)}$ at a \he4 density of
$\rho=0.022\,$\AA$^{-3}$ which is close to the experimental
equilibrium density. This function most closely corresponds to the
pair wave function $\psi(r)$ shown in Fig. \ref{fig:psiplot}.
Unlike the pair wave functions shown in Fig. \ref{fig:psiplot}, the
corresponding quantities $\sqrt{g_{\rm IB}(r)}$ show, with increasing
mass, an increasing nearest neighbor peak which is due to the
short-ranged structure of the host liquid. On the other hand, the
correlation hole of the muonium becomes very large although not
quite as pronounced as that of the vacuum system. The feature that
stands out again is the large overlap between the short-ranged
wave function and the repulsive core of the interaction.

\begin{figure}[H]
  \centerline{\includegraphics[width=0.3\textwidth,angle=270]{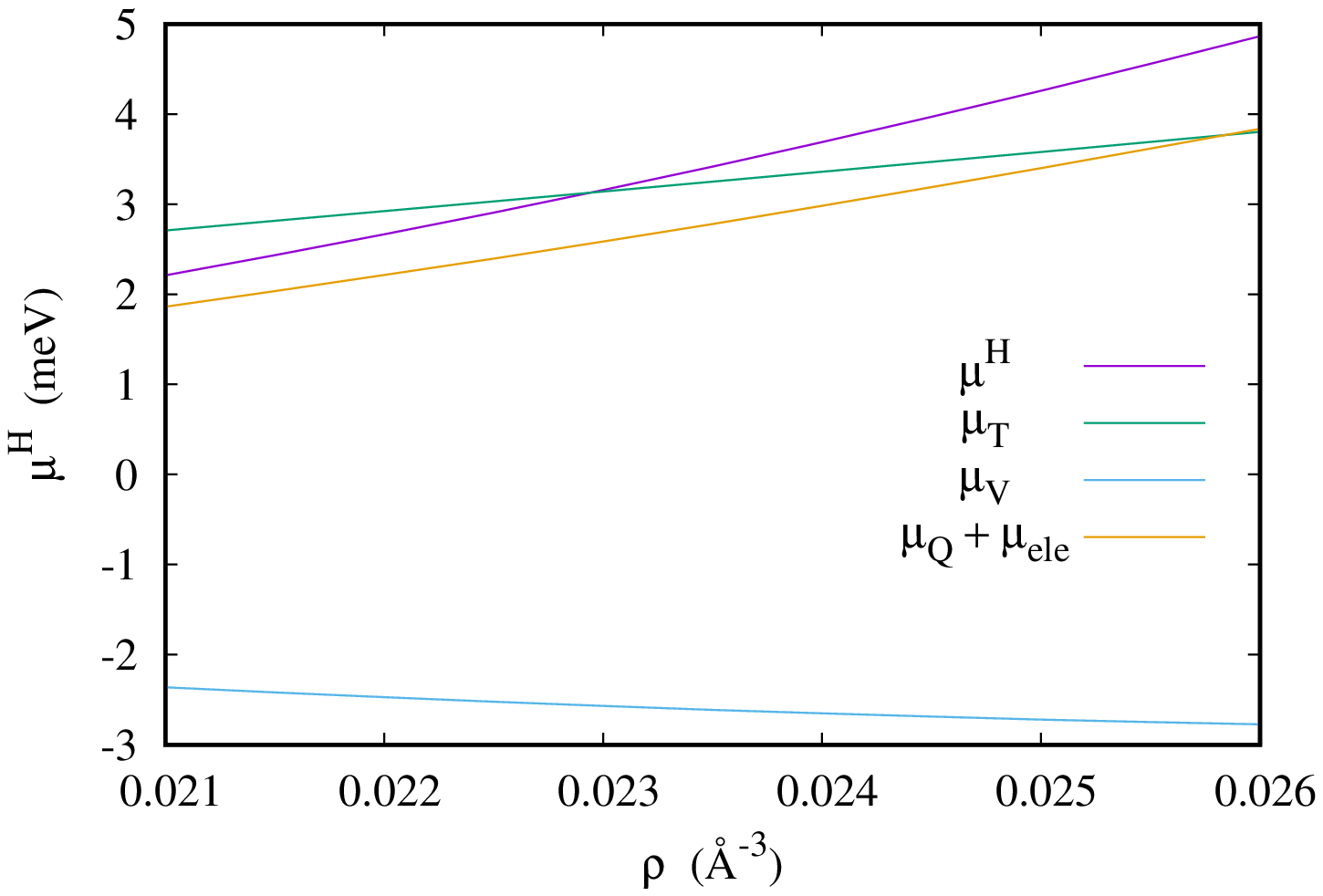}}

  \centerline{\includegraphics[width=0.3\textwidth,angle=270]{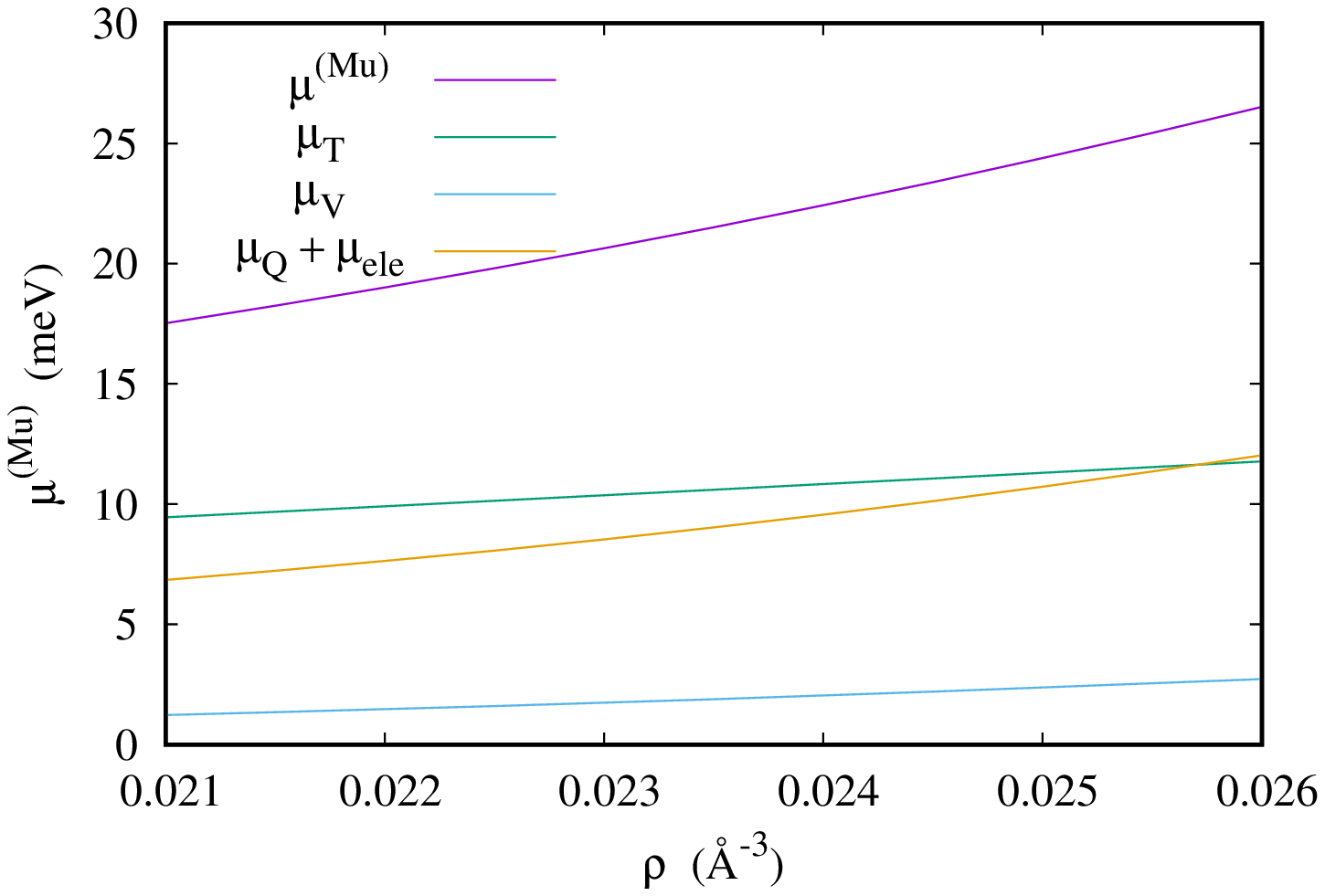}}

  \caption{(color online) The figures show the individual contributions
    to the chemical potential of hydrogen impurities (top panel)
    and muonium impurities (bottom panel) as spelled out in Eqs.
    \eqref{eq:muImpu}, \eqref{eq:eRImpu}, and \eqref{eq:eQImpu}.
    \label{fig:eosparts}}
  \end{figure}

Interesting information about the energetics of the systems is
obtained by comparing the different parts of the
energy. Figs. \ref{fig:eosparts} show, for hydrogen and muonium atoms,
the individual parts of the chemical potential: $\mu_V^{(I)}$ is the
potential energy, $\mu_T^{(I)}$ the cost of kinetic energy due to the
bending of the wave function at short distances, and
$\mu_Q^{(I)}+\mu^{(I)}_{\rm ele}$ the contribution from many-body
correlations.  While the energy scale is rather different, the
relative size of the individual contributions is comparable. There is,
however, one remarkable difference: The potential energy contribution
$\mu_V$ is {\em negative\/} for hydrogen atoms but {\em positive\/}
for muonium.  The reason for this difference is the overlap of the
pair distribution function with the short-ranged part of the
interaction as demonstrated in Figs. \ref{fig:psiplot} and
\ref{fig:gibplot}.

\begin{figure*}[t]
  \centerline{\includegraphics[width=0.5\textwidth,angle=270]{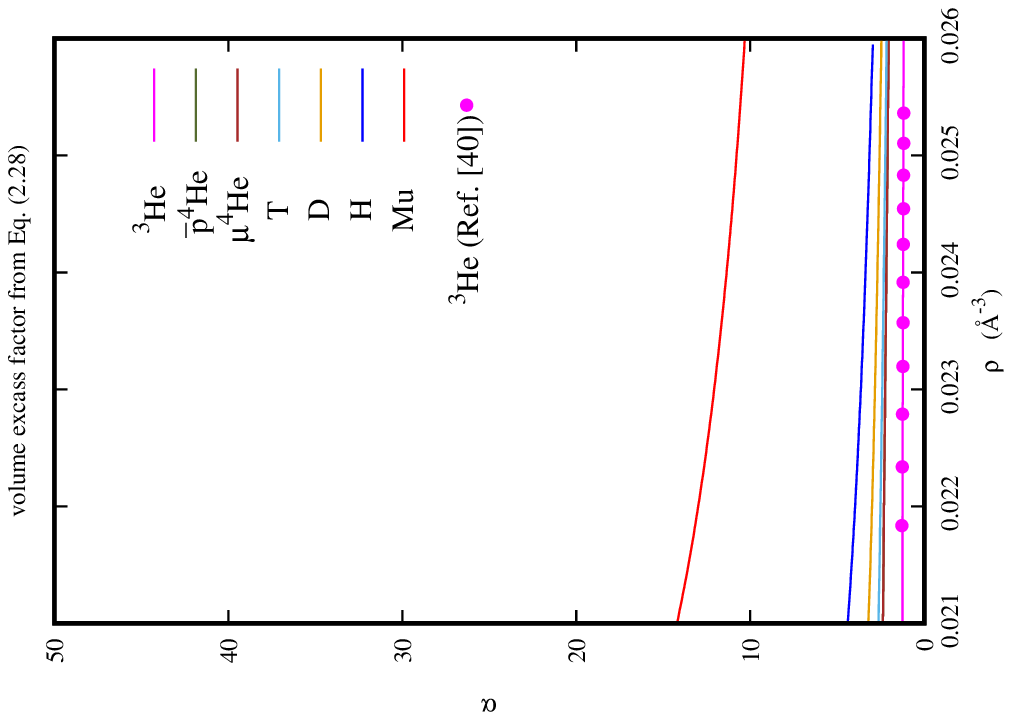}
    \includegraphics[width=0.5\textwidth,angle=270]{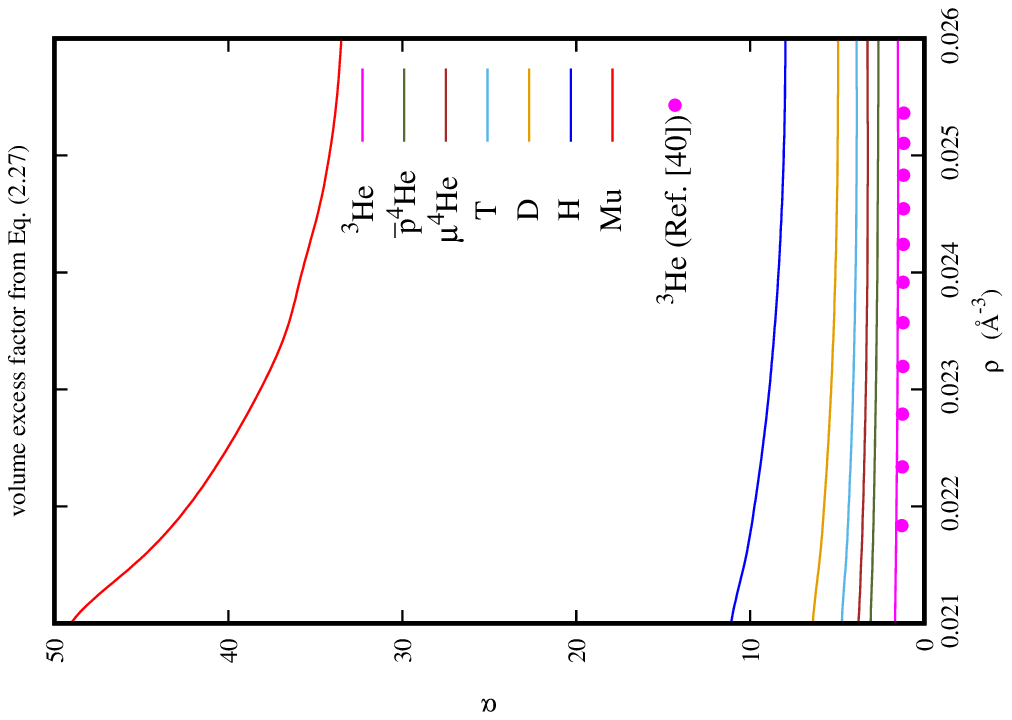}}
  \caption{(color online) The left figure shows the volume excess
    factor $\alpha$ as obtained from Eq. \eqref{eq:alphaEOS} by a fit
    of the density-dependence of the chemical potential, and the right
    figure shows the same quantity as obtained from the
    long-wavelength limit \eqref{eq:alphaSofq} of the static structure
    function.\label{fig:alpha}}
\end{figure*}

As pointed out above, the comparison of different expression
\eqref{eq:alphaSofq} and \eqref{eq:alphaEOS} for the volume excess
factor $\alpha$ gives a very conservative estimate of the importance
of higher-order elementary diagrams and multiparticle correlation
functions.  Figs. \ref{fig:alpha} shows these quantities obtained in
the two different ways; for the calculation of the quantity
\eqref{eq:alphaEOS} we have taken the compressibility from the
derivative \eqref{eq:mceos} obtained by a numerical fit of the
equation of state of bulk \he4. In agreement with earlier work
\cite{SKJLTP} one concludes that the convergence of the set of
elementary diagrams is similar for H, D, T, \he4, $\mu$\he4, and
$\bar{\rm p}^4$He, but considerably worse for muonium. This is hardly a
surprise considering the much larger correlation hole shown in
Fig. \ref{fig:gibplot}.

\subsection{Dynamics}
Throughout this work we have used the renormalized form of the
self-energy. It includes a larger class of contributions and provides,
as the discussion around Eqs. \eqref{eq:sigmaGWons} and
\eqref{eq:Istar} show, a more consistent treatment of the
singe-particle propagator.

The quantity of predominant interest in bulk \he4 is the response of
the bulk liquid to neutron scattering described by the dynamic
structure function $S(k,\omega)$.  Here we are, instead, interested in
the motion of single atoms within the host liquid which is characterized
by the self-energy $\Sigma^{(I)}(k,\omega)$. At low energies,
the impurity moves freely through the superfluid; this is simply
because the dispersion relation of the impurity is quadratic in the
momentum, see Eq. \eqref{eq:epslong} whereas the phonon dispersion
relation is linear, hence the energy of the impurity is less that the
lowest excitation of the host liquid. The only many-body effect is
hydrodymanic backflow \cite{FeynmanBackflow} which has the consequence
that the particle acquires an effective mass.

\begin{figure}[H]
  \centerline{\includegraphics[width=0.35\textwidth,angle=270]{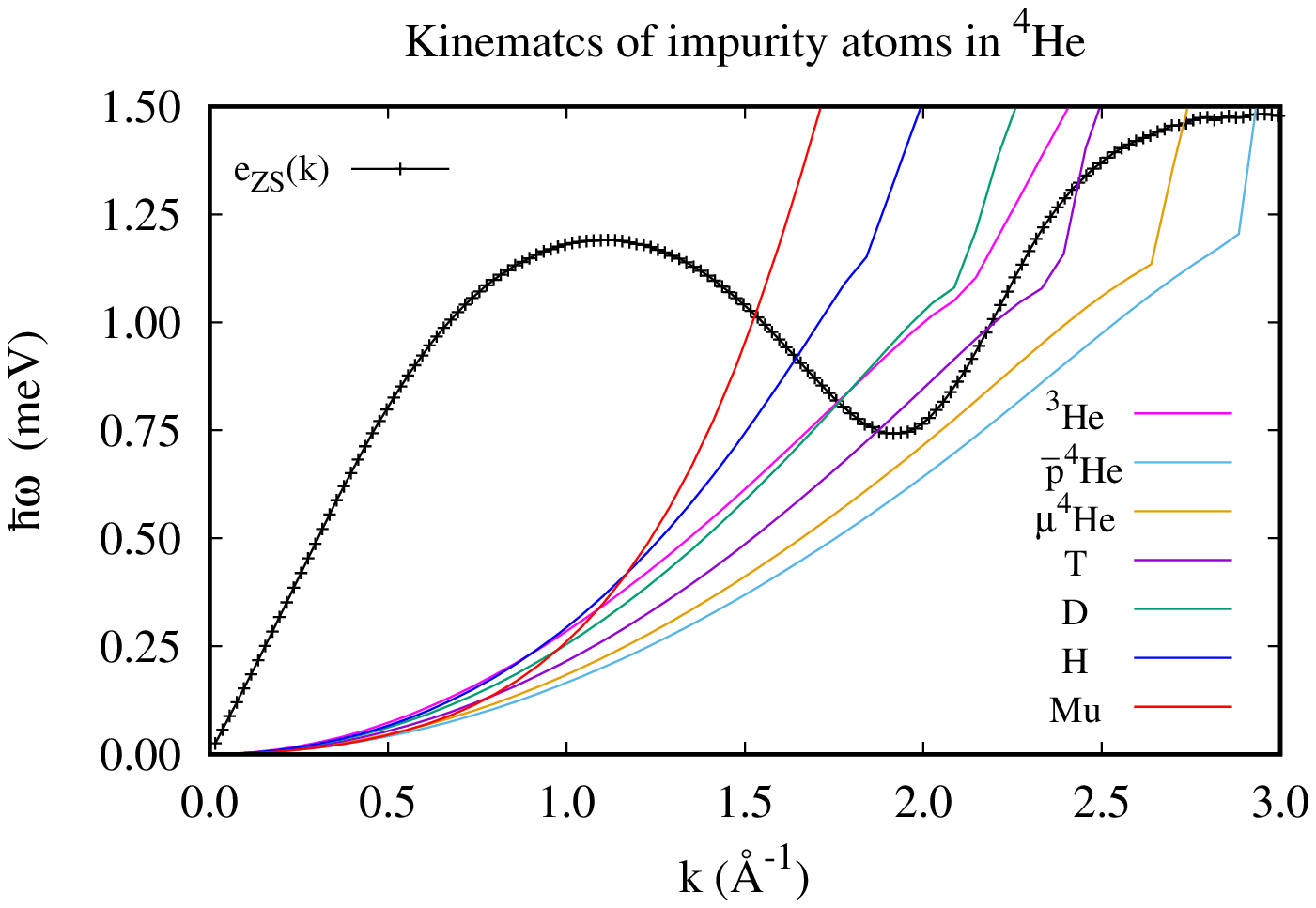}}
  \caption{(color online) The figure shows the calculated phonon-roton
    spectrum from Ref. \onlinecite{He4Dispersion} (crosses) and the
    single particle spectra for the impurity atoms considered in this
    work as indicated in the legend.\label{fig:kinematics}}
\end{figure}

At higher momentum transfers the impurity atom can couple to the
excitations of the host \he4 liquid and the impurity motion gets
damped.  The kinematic situation is shown in
Fig. \ref{fig:kinematics}.
We stress that the notion that damping occurs at the point where
the impurity dispersion relation intersects with the phonon-roton spectrum
is valid valid only when the self-energies in the denominator
of Eq. \eqref{eq:sigmau} are exhausted by the impurity dispersion
relation and the phonon-roton spectrum. This is only the case at
short wave length but not at the wave length where the crossing
actually occurs; we will get back to this point shortly.

A number of interesting observations apply: Apparently the spectrum of
a muonium atom is quadratic only for momentum transfers less than
$0.3\,$\AA$^{-1}$. For higher momenta it bends, unlike the spectrum of
all other impurities, strongly upwards which leads to the smallest
momentum of all impurity atoms considered here in which the particle
can move freely. This is evidently also a consequence of the much
larger effective size of the particle.

On the other side of the mass range we see that the spectrum of muonic
\he4 and antiprotonic \he4 pass under the roton minimum and can,
hence, move freely up to wave number of almost $k\approx
3\,$\AA$^{-1}$ corresponding to a velocity of almost 500 m/sec.
At larger wave numbers the atom can couple to the Pitaevskii plateau.

\begin{figure}[H]
  \centerline{\includegraphics[width=0.35\textwidth,angle=270]{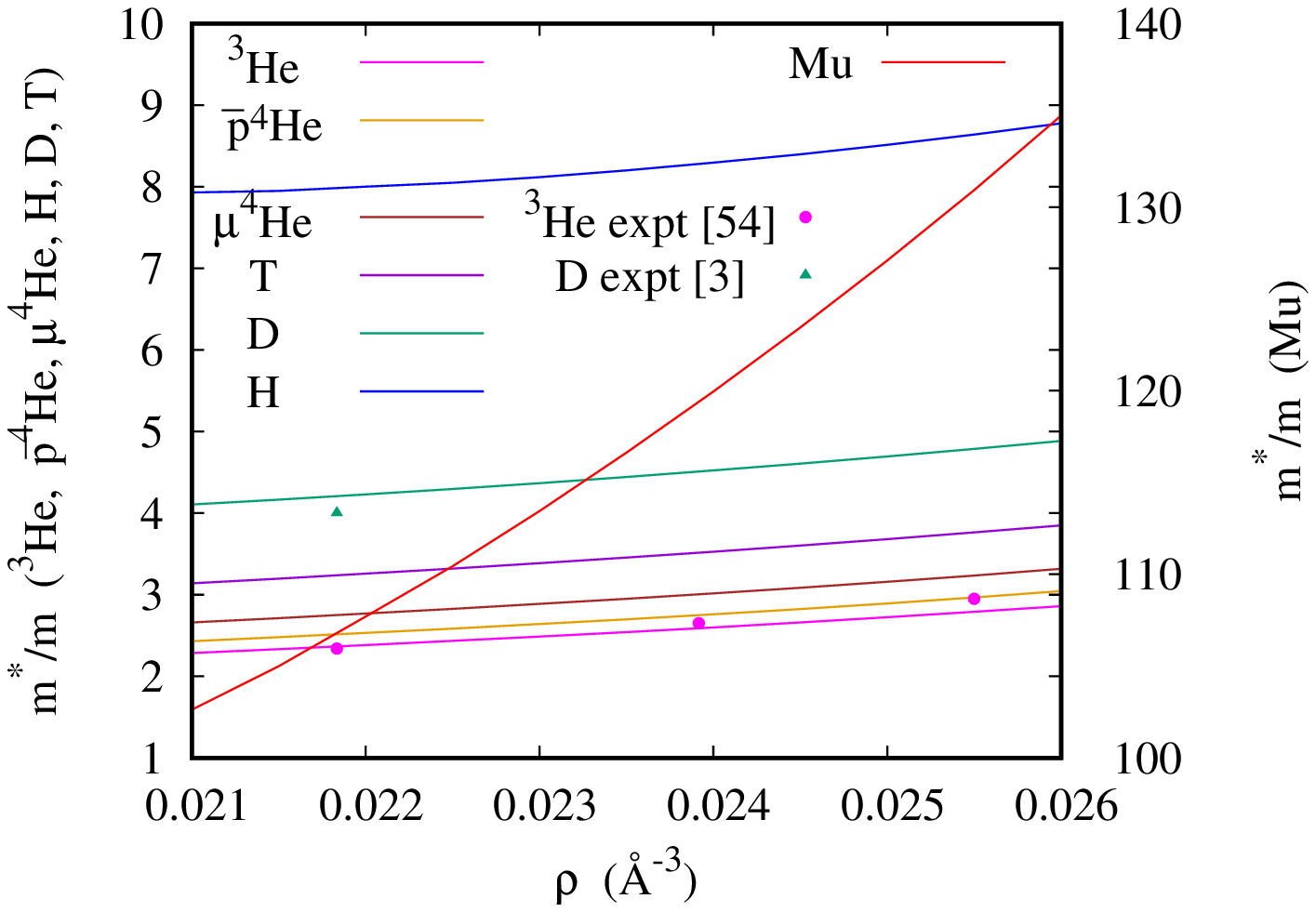}}
  \caption{(color online) The figure shows the calculated effective
    mass for the impurity atoms considered in this work. The muonium
    effective mass is given on the right scale of the figure.
    \label{fig:massplot}}
\end{figure}

\begin{figure*}[t]
  \centerline{\includegraphics[width=0.35\textwidth,angle=270]{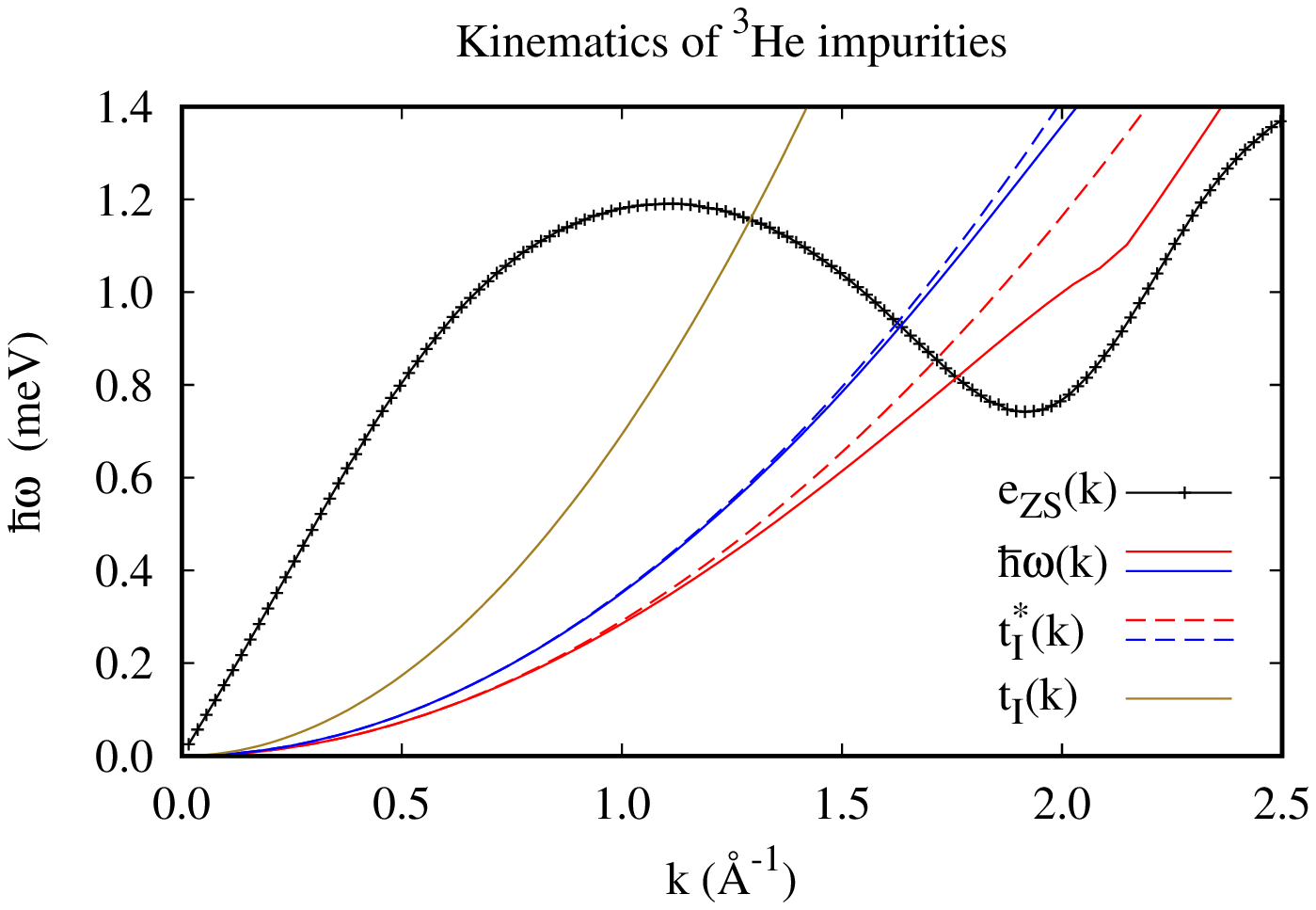}
    \includegraphics[width=0.35\textwidth,angle=270]{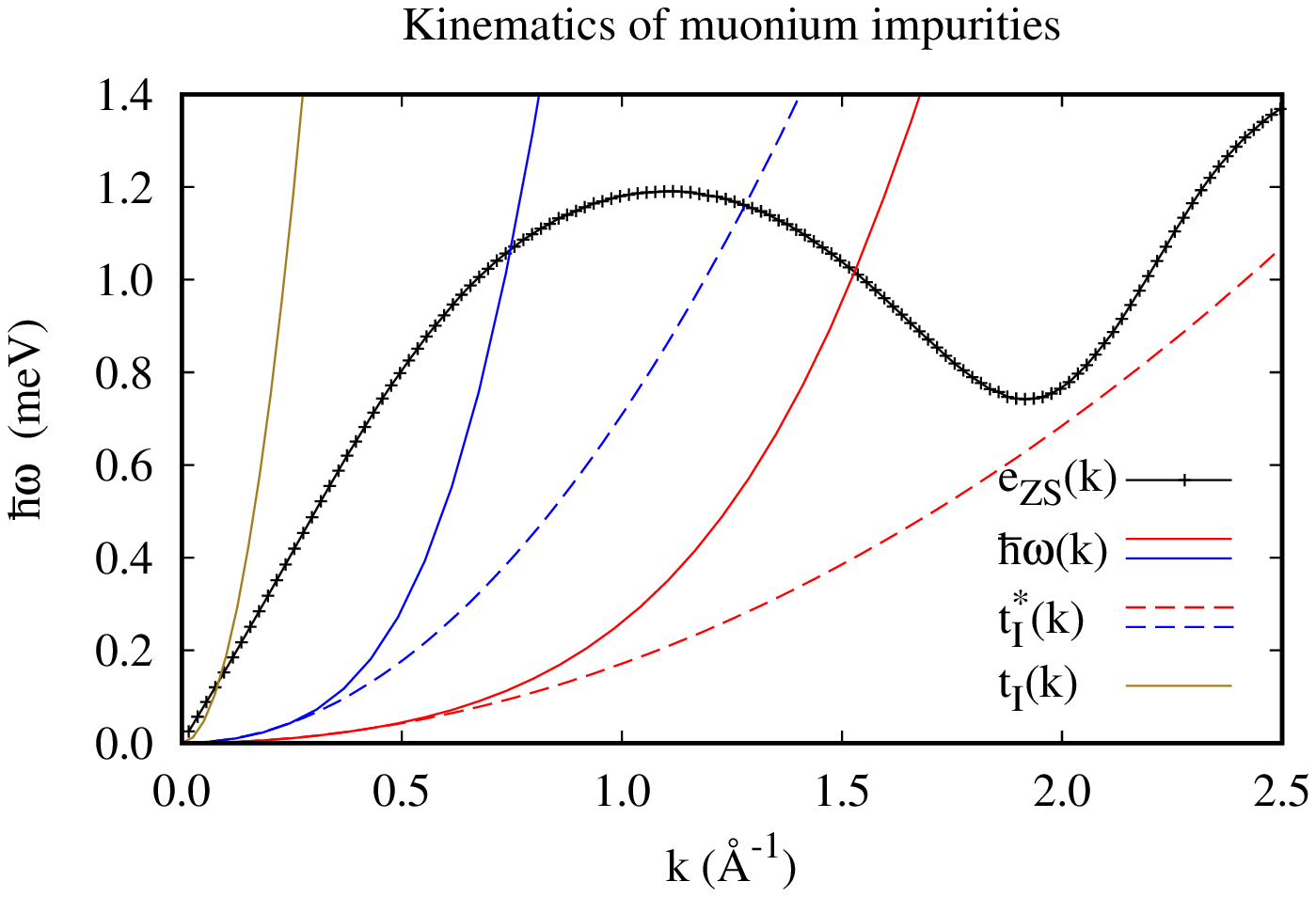}}
  \caption{(color online) The left figure shows the dispersion
    relation of a \he3 impurity using the full self-energy (red solid
    line) and using the G0W approximation \eqref{eq:sigmau2} (blue
    solid line) along with a quadratic fit $t^*(k) = \hbar^2
    k^2/2m_I^*$ (dashed red and blue lines) to the dispersion
    relation. Also shown is, for reference, the spectrum of a free
    particle $t(k) = \hbar^2 k^2/2m_I$.  The right figure gives the
    same information for a muonium
    impurity.\label{fig:MuHeKinematics}}
\end{figure*}

It is common to characterize the motion of single particles by an
effective mass as we have done above, see Eqs. \eqref{eq:Idef} and
\eqref{eq:muren}. We have compared the results of the direct
calculation with a fit to the spectrum \eqref{eq:Dispersion} and
verified the agreement within numerical accuracy. Our results are
shown, as a function of density, in Fig. \ref{fig:massplot}.  We
stress again that an effective mass formula is valid for muomium only
for rather small wave numbers, see Fig. \ref{fig:MuHeKinematics}
whereas it is a good approximation for all other particles considered
here up to the point where the particle can couple to phonon-roton
excitations. The most dramatic change occurs, of course,  again
for muonium where the effective mass ratio becomes over 100.

As a technical point we note that the dispersion relations are, in the
regime where the particle can move freely through the medium, almost
identical to when we omit the full background self-energy
$\Sigma^{(B)}(k,\omega)$. We show these features in
Fig. \ref{fig:MuHeKinematics}. Of course, damping begins at a much
lower energy.

\begin{figure}[H]
  \centerline{\includegraphics[width=0.35\textwidth,angle=270]{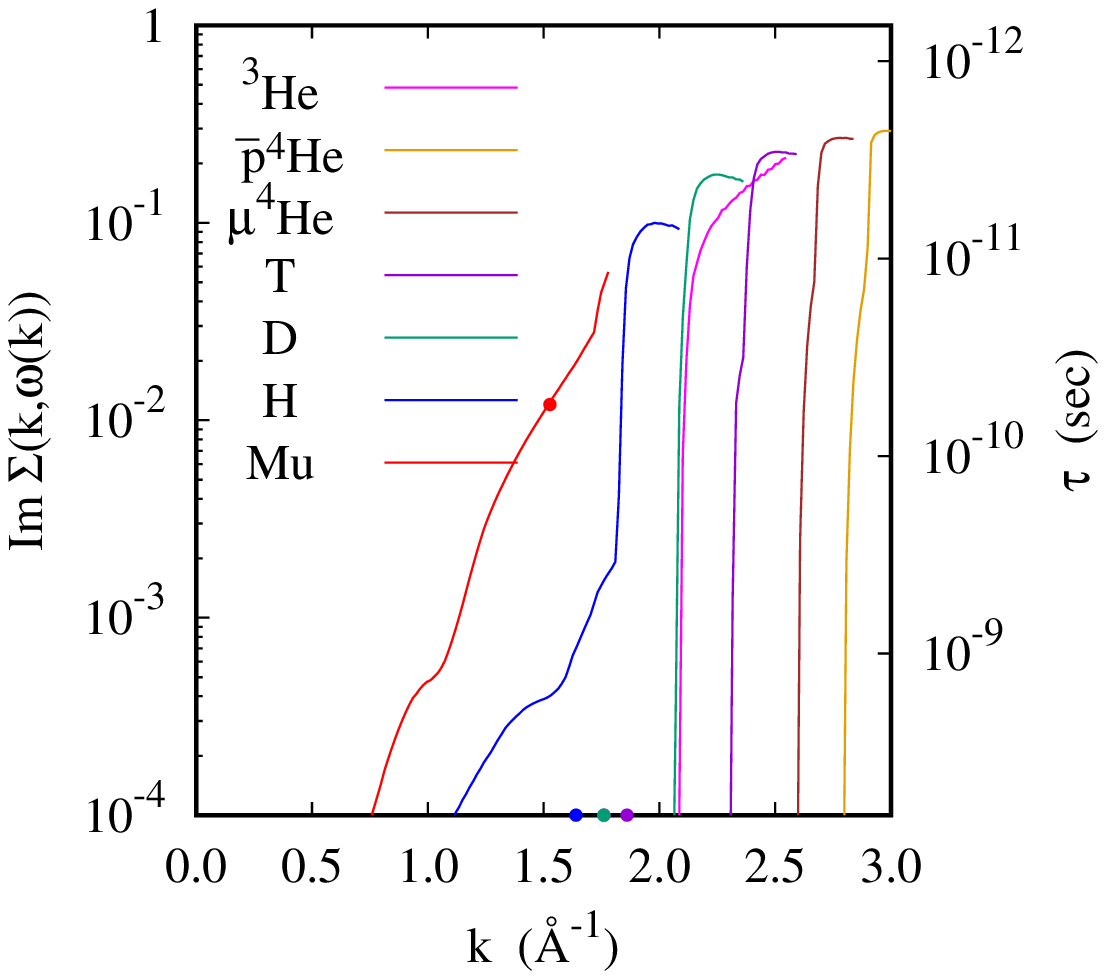}}
  \caption{(color online) The figure shows at a density of $\rho=
    0.022\,$\AA$^{-1}$ the imaginary part of the self-energy (left
    scale) and the resulting lifetime \eqref{eq:tau} (right
    scale). The colored dots show the points where the corresponding
    dispersion relation $\hbar\omega(k)$ crosses the phonon-roton
    spectrum. The dispersion relations of $\mu$\he4 and $\bar{\rm p}$\he4
    does not cross the phonon-roton spectrum on the scale of the
    plot. For \he3, H, D, and T we have moved the point to $10^{-4}$
    to make it visible in the plot, the crossing actually occurs
    where the imaginary part of the self-energy is zero.  The points
    for \he3 and D are on top of each other.\label{fig:lifetimes}}
\end{figure}

Let us now turn to the damping of the impurity
motion. Fig. \ref{fig:lifetimes} shows the imaginary part of the
self-energy for the impurity atoms under investigation here which
determines the lifetime of the single particle excitations through
\begin{equation}
  \tau(k) = \hbar\, \left|\Im\Sigma(k,\omega(k))\right|^{-1}\,.
\label{eq:tau}
\end{equation}

The lifetime also determines the mean free path
\begin{equation}
  d(k) = v_g(k)\tau(k)\label{eq:mfp}\,.
\end{equation}
where $v_q = d\omega(k)/dk$ is the group velocity.  In the areas shown
in Fig. \ref{fig:lifetimes} where the single particle spectra acquire
a finite imaginary part, the mean free path drops quickly to a few
\AA.

\begin{figure*}[t]
  \centerline{\includegraphics[width=0.35\textwidth,angle=270]{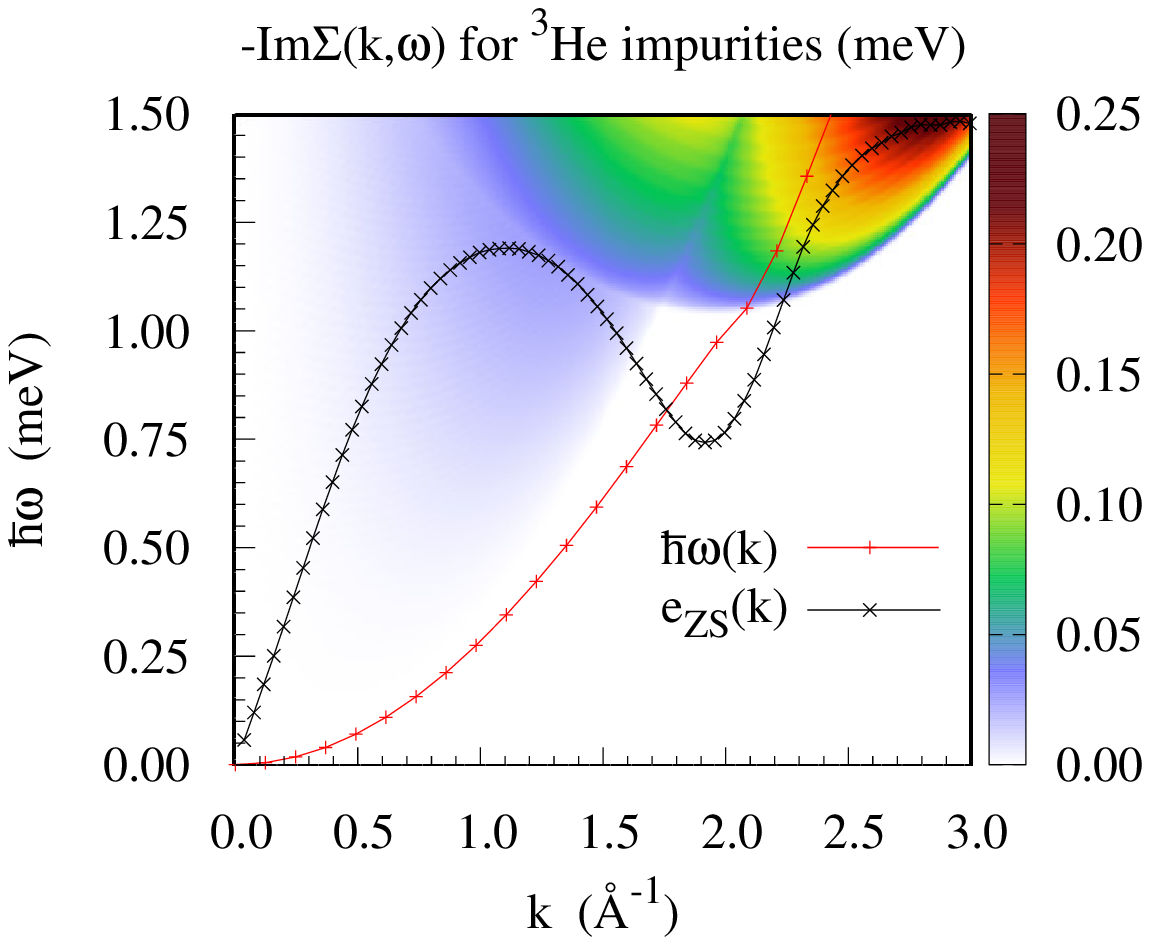}
    \includegraphics[width=0.35\textwidth,angle=270]{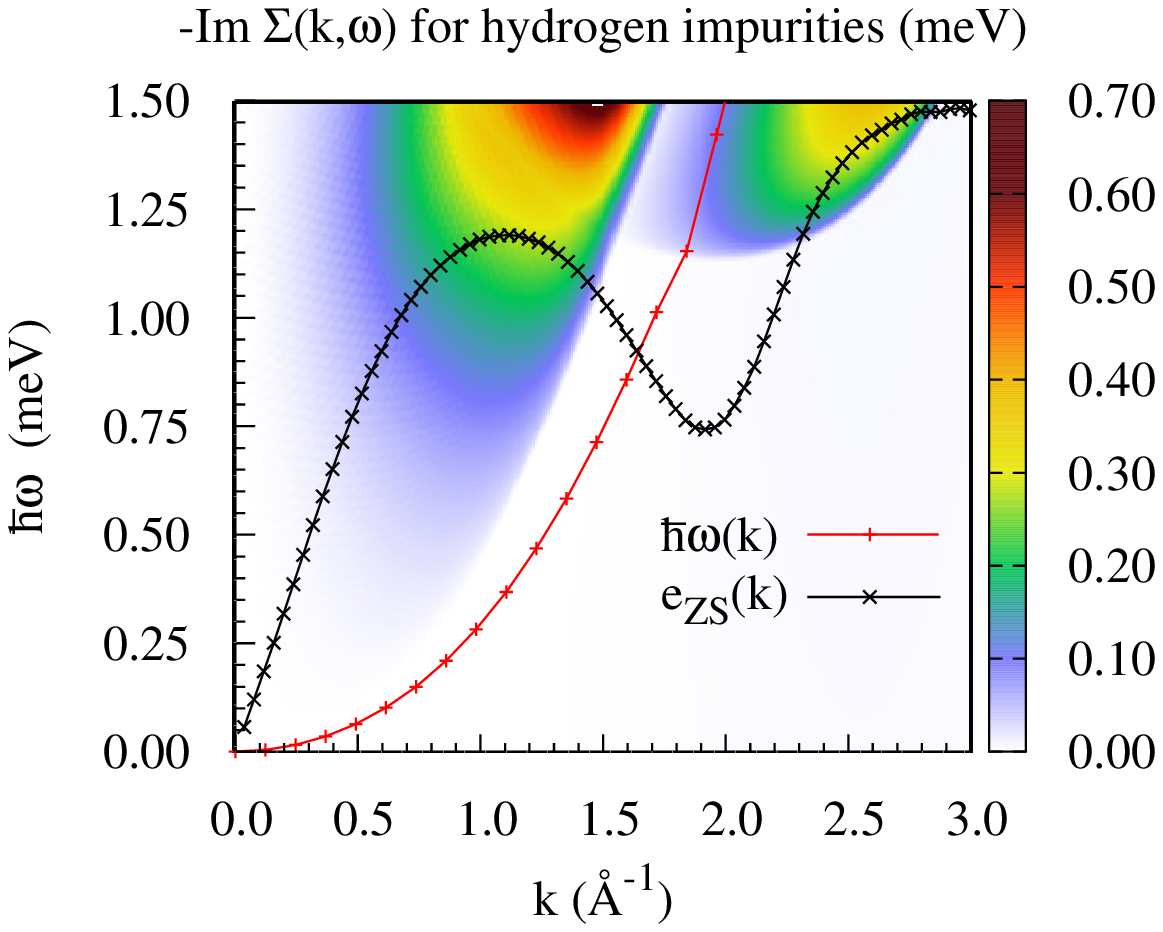}}
  \centerline{\includegraphics[width=0.35\textwidth,angle=270]{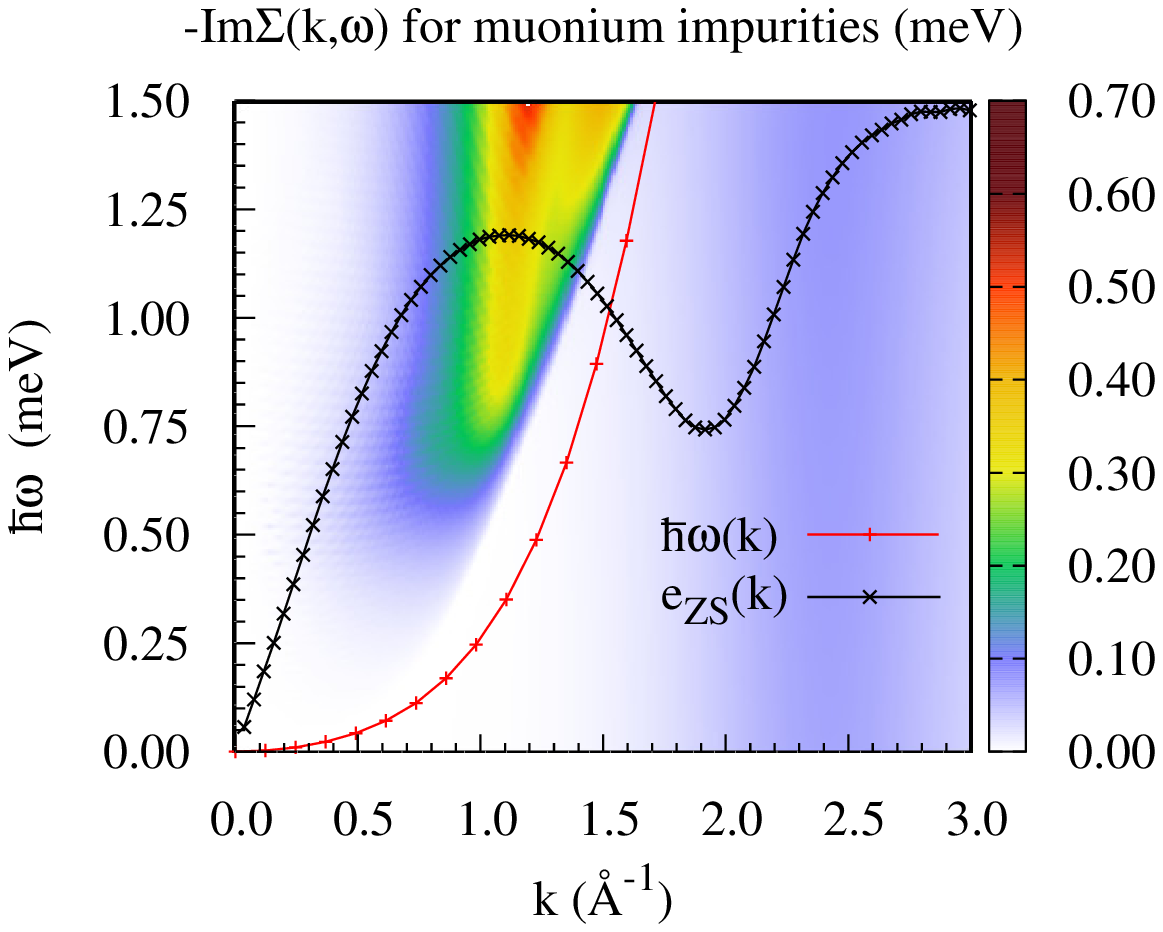}
    \includegraphics[width=0.35\textwidth,angle=270]{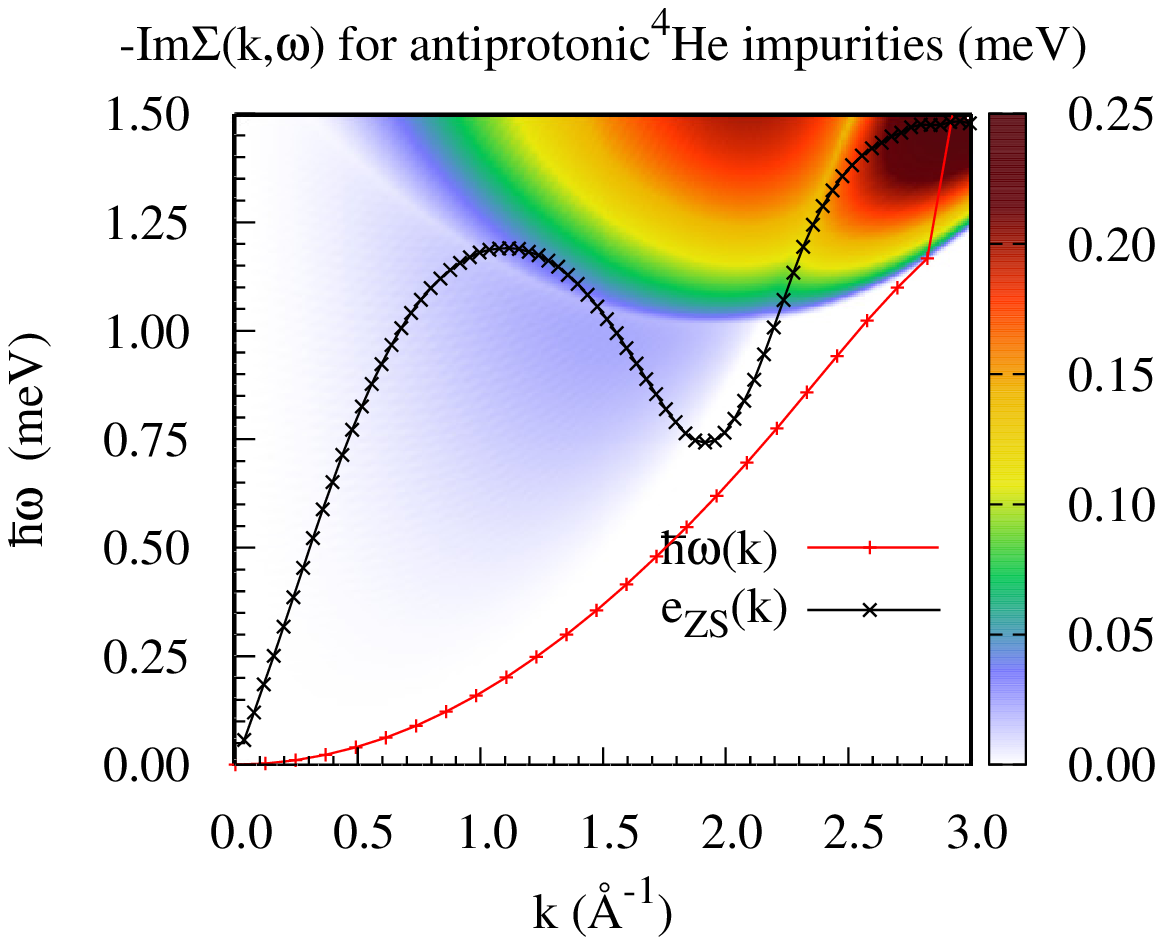}}
  \caption{(color online) The figures shows color maps of the
    imaginary part of the self-energy for four impurity atoms at a
    \he4 density of $\rho=0.022\,$\AA$^{-3}$. Also shown is the
    phonon-roton spectrum of the host liquid \he4 (black lines) and
    the impurity dispersion relation. The particles can move freely
    where $\Im \Sigma(k,\omega)$ is zero which is indicated by the
    white area.
   \label{fig:ImSig}}
\end{figure*}

The above-mentioned feature that the onset damping does not occur
at the point of intersection of the impurity dispersion relation
and the phonon-roton spectrum is already clear from the expression
\eqref{eq:sigmau}; the details depend not only on the specific features
of the self-energy, but also on the kinematic constraint of energy and
momentum conservation. We show in figs. \ref{fig:ImSig} four
cases: \he3, H, Mu, and $\bar{\rm p}$\he4. 

We have chosen these cases because they exemplify typical
configurations: \he3 impurities because \he3-\he4 mixtures are perhaps
the best studied system
\cite{EbnE,EdwR,BAP,OuY,BBP,HsuPines,HsuPiAl,MixMonster}, H for
comparison with muonium, and $\bar{\rm p}$\he4 because the dispersion
relation passes under the roton minimum.  All of the cases show a
remarkable amount of structure which is due to the coupling of
different perturbations of the host liquid and the impurity. We have
studied these effects in much detail for pure \he4 in
Ref. \onlinecite{He4Dispersion} and have verified that the theoretical
description reproduces even fine details of the experimental findings.
The results for \he3 are experimentally accessible by neutron
scattering \cite{Fak,1993-Scherm-mixtures,1994-Fak-Scherm-mixtures}
but the accuracy achieved at the time these experiments were made does
not reach the one for pure \he4 reported in
Ref. \onlinecite{He4Dispersion} to allow a detailed comparisons with our
results.  Experimental access to the off-shell dynamics of hydrogen
isotopes is much more difficult, we refrain there here from a close
analysis of the individual features of these systems.

\section{Summary}
We have in this paper presented the extension of our earlier
development of dynamic many-body theory (DMBT) \cite{eomI,eomIII}
to the motion of impurity atoms in a dense quantum fluid, specifically
\he4.  The studies were motivated by recent experiments on the
generation of high intensity, superthermal muonium beams for gravity
and laser spectroscopy experiments\cite{Muonium}. We have verified our
theoretical methods by comparison with experimental data as far as
available \cite{EbnE,Reynolds91,Greywall79,Hayden95} but note, of
course, that muonium impurities probe a rather different physical
situation.

Our work raises a number of interesting questions. One is, of course,
the energetics of the system. From our knowledge about the chemical
potential of H, D, and T atoms one would conclude that the chemical
potential of muonium is much higher. The large correlation hole
generated by the muonium challenges, of course, the convergence of
diagrammatic many-body methods; a Monte Carlo evaluation of the
chemical potential similar to what we did for Magnesium impurities
\cite{MgDimers} would eliminate this uncertainty. However, such a
calculation does not eliminate the problem that the wave function
probes the interaction at distances of the order of half of the core
size. At such short distances, the interaction is purely
phenomenological and not well known; in fact it is not even clear if
the concept of a static potential is valid.

Being aware of these concerns, we have derived the equations of motion
for the impurity dynamics.  The forms \eqref{eq:sigmau} and
\eqref{eq:sigmar} describe the very simplest version of the
multiparticle dynamics which is outlined in detail in the supplemental
material \cite{eom4suppl}. The theory was very successful in bulk \he4
describing even fine details of the dynamic structure function
\cite{He4Dispersion}. While some of the approximations going into the
form \eqref{eq:sigmar} can certainly be removed, the expressions used
here describe the relevant effects, namely the self-consistency of the
impurity self-energy on the left and the right hand side of Eq.
\eqref{eq:sigmau} and the coupling of the impurity motion to the
physical phonon roton spectrum. The latter is, in the form
\eqref{eq:sigmau2}, included only in the Feynman approximation which
predicts a phonon-roton spectrum that is about a factor of 2 too
high. Extensions of the method can be envisioned if and when
sufficiently accurate experimental data on the self-energy in the full
complex plane become available. A particularly promising project would
be to carry out neutron scattering experiments on \he3-\he4 mixtures
of the kind described in Refs.
\onlinecite{Fak,1993-Scherm-mixtures,1994-Fak-Scherm-mixtures} but
with an accuracy that comparable to what was achiieved for pure \he4
\cite{He4Dispersion}.

None of these considerations address the obvious discrepancy between
theoretical predictions of the muonium chemical potential inside the
\he4 and the observed low energy and sharp velocity distribution of
the exiting particles \cite{Muonium}. A conceivable mechanism is the
coupling to ripplon excitations during the transition through the \he4
surface. We have studied the effect of ripplon emission by impending
hydrogen atoms from the vacuum onto the \he4 surface \cite{hscatt}, in
that case the effect of ripplon emission is large albeit the effect is
suppressed by the mechanism of quantum reflection \cite{Brenig80}.
Work in this direction is under way.

\begin{acknowledgments}
  The author would like to thank Anna S\'ot\'er, David Taqqu, and Klaus
  Kirch for bringing his attention to the muonium problem and for
  communicating their results and ideas prior to publication. Thanks
  are also due to Robert Zillich for discussions.
\end{acknowledgments}

\bibliography {papers}
\bibliographystyle{apsrev4-1}
\end{document}